\documentclass[aps,prd,twocolumn,preprintnumbers,showpacs,nofootinbib,amssymb]{revtex4}
\usepackage{graphicx}
\usepackage{amsmath}
\usepackage{amssymb}
\usepackage{amsfonts}
\usepackage{bm}
\usepackage{color}

\def\be{\begin{equation}}
\def\ee{\end{equation}}
\def\bea{\begin{eqnarray}}
\def\eea{\end{eqnarray}}

\begin{document}

\title{Core mass -- halo mass relation of 
bosonic and fermionic dark matter halos \\
 harbouring a supermassive black hole}
\author{Pierre-Henri Chavanis}
\email{chavanis@irsamc.ups-tlse.fr}
\affiliation{Laboratoire de Physique Th\'eorique, Universit\'e de Toulouse,
CNRS, UPS, France}

\begin{abstract}

We study the core mass -- halo mass relation of bosonic dark matter
halos, in the form of self-gravitating Bose-Einstein condensates, harbouring a
supermassive black hole. We use the ``velocity dispersion tracing'' relation
according to which
the velocity dispersion in the core $v_c^2\sim GM_c/R_c$ is of the same order as
the velocity dispersion in the halo $v_h^2\sim GM_h/r_h$ (this relation can
be justified from thermodynamical arguments) and the approximate analytical
mass-radius relation of the quantum core in the presence of a central black hole
obtained in our previous paper [P.H. Chavanis, Eur. Phys. J. Plus
{\bf 134}, 352 (2019)]. For a given minimum halo mass
$(M_h)_{\rm min}\sim 10^8\, M_{\odot}$ determined by the observations, the only
free parameter of our model is the scattering length $a_s$ of the bosons (their
mass $m$ is then determined by the characteristics of the minimum halo). For
noninteracting bosons and
for bosons with a repulsive self-interaction, we find that the core mass $M_c$
increases with the halo mass $M_h$ and achieves a maximum value $(M_c)_{\rm
max}$ at some halo mass 
$(M_h)_{*}$ before decreasing. The whole series of equilibria is stable. For 
bosons with an attractive self-interaction, we find
that the core mass achieves a maximum value $(M_c)_{\rm max}$ at some halo mass 
$(M_h)_{*}$ before decreasing. The series of equilibria becomes
unstable above a maximum
halo mass  $(M_h)_{\rm max}\ge (M_h)_{*}$. In the absence of black hole
$(M_h)_{\rm max}=(M_h)_{*}$. At that point, the quantum
core (similar to a dilute axion star) collapses.  We
perform a similar study for fermionic dark matter halos. We find that they
behave similarly to bosonic dark matter halos with a repulsive self-interaction,
the Pauli principle
for fermions playing the role of the 
repulsive self-interaction for bosons.

\end{abstract}


\maketitle

\section{Introduction}

The nature of dark matter (DM) is still unknown and remains one of the greatest
mysteries of modern cosmology. The standard cold dark matter (CDM)
model, assuming the existence (still hypothetical) of weakly interacting massive
particles
(WIMPs) of mass $m\sim {\rm
GeV/c^2}$, works remarkably well at large (cosmological) scales
\cite{planck2013,planck2016} but  encounters serious
problems at small (galactic) scales that are referred to as the ``cusp-core
problem'' \cite{nfw,observations}, the ``missing satellite problem''
\cite{satellites}, and the ``too big to fail problem'' \cite{toobigtofail}. The
expression ``small-scale crisis of CDM''
has been coined. In order to solve these problems, some authors have
proposed to take  the quantum nature
of the DM particle into account (see an exhausive list of references in 
Refs. \cite{prd1,modeldm} and in the reviews
\cite{srm,rds,chavanisbook,marshrevue,leerevue,braatenrevue}). Indeed, quantum
mechanics creates an effective
pressure even at zero thermodynamic temperature ($T_{\rm th}=0$) that may
balance the gravitational
attraction at small
scales and lead to cores instead of cusps. The DM particle
could be a fermion, like a massive neutrino, or 
a boson in the form of a Bose-Einstein condensate (BEC), like an ultralight
axion. In order to match the
characteristics of the smallest halos like dwarf spheroidals (dSphs) that are
interpreted as purely quantum objects
(see below), the mass of the fermion should be of the order of $m\sim 170\, {\rm
eV/c^2}$ while the mass of the boson should be of the order of $m\sim 2.92\times
10^{-22}\, {\rm eV/c^2}$ if it has a vanishing self-interaction, or lie in the
range
$2.19\times 10^{-22}\, {\rm eV}/c^2<m<2.92\times 10^{-22}\, {\rm
eV}/c^2$ if it has an attractive self-interaction, and in the
range $2.92\times 10^{-22}\, {\rm eV}/c^2<m<1.10\times 10^{-3}\, {\rm
eV}/c^2$ if it has a repulsive self-interaction (see Appendix D of
\cite{suarezchavanis3} and Ref. \cite{mcmh}).\footnote{These values are 
indicative. They are orders of magnitude that could be improved by making
a more detailed
comparison with observations. The arguments of \cite{suarezchavanis3,mcmh}
suggest that the mass of a boson with an attractive self-interaction
should be slightly smaller than the mass of a noninteracting boson
($m\sim 2.92\times 10^{-22}\, {\rm eV/c^2}$), while
the mass of a boson with a repulsive self-interaction can  be  up
to $20$ orders of magnitude larger than the mass of a noninteracting boson. As
discussed in Appendix D.4 of \cite{suarezchavanis3}, a mass larger than
$2.92\times 10^{-22}\, {\rm eV/c^2}$ could alleviate some
tensions with the observations of the Lyman-$\alpha$ forest
encountered in the noninteracting model. As a result, a
repulsive self-interaction ($a_s>0$) is priviledged over an attractive
self-interaction ($a_s<0$). A repulsive self-interaction is also favored by
cosmological constraints \cite{shapiro,suarezchavanis3}. In this respect, we
recall that
theoretical models of particle physics  usually lead to particles with
an attractive self-interaction (e.g., the QCD axion). However, some authors
\cite{fan,reig} have pointed out the possible existence of particles with a
repulsive self-interaction (e.g., the light majoron).}

In these quantum models, sufficiently large DM halos have a ``core-halo''
structure which results
from a process of violent collisionless relaxation \cite{lb} and gravitational
cooling \cite{seidel94,gul0,gul}. This core-halo structure has
been evidenced in direct numerical simulations of noninteracting BECDM
\cite{ch2,ch3,moczprl,veltmaat,mocz}. The quantum core stems from the
equilibrium
between the quantum pressure and the gravitational attraction (ground
state).\footnote{The quantum pressure arises from the Pauli exclusion principle
for fermions and from the Heisenberg uncertainty principle for bosons. A
repulsive self-interaction may also help stabilizing the quantum core. By
contrast, an attractive  self-interaction tends to destabilize the quantum core
above a maximum mass $M_{\rm max}=1.012\, \hbar/\sqrt{Gm|a_s|}$
identified in \cite{prd1}.}
Quantum
mechanics stabilizes DM halos against gravitational collapse, leading to flat
cores instead of cusps. The core mass-radius
relation
$M_c(R_c)$ can be obtained numerically by solving the differential equation of
quantum hydrostatic equilibrium \cite{prd2}, or
approximately by making a Gaussian ansatz
for the density profile \cite{prd1}. 
On the other hand, the halo (atmosphere) is relatively independent of quantum
effects. It is
similar to the Navarro-Frenk-White (NFW)
profile \cite{nfw} produced in CDM simulations or to the empirical Burkert
profile
\cite{observations} deduced from the observations.\footnote{In the
case of BECDM, the halo is due to quantum interferences of excited
states. It is made of granules (quasiparticles) of the size  of the solitonic
core $\lambda_{\rm dB}\sim \hbar/mv\sim 1\, {\rm kpc}$ (de Broglie wavelength)
and of effective mass $m_*\sim\rho\lambda_{\rm dB}^3\sim 10^7\, M_{\odot} \gg m$
\cite{ch2,ch3,hui,bft}. These supermassive
granules can cause the ``collisional''
relaxation of the halo on a timescale smaller than the age of the universe (by
contrast,
the relaxation due to the ultralight particles of mass $m$ would be completely
negligible).} It is responsible for the
flat rotation curves of the
galaxies at large distances. We shall approximate this halo by an isothermal
sphere with an effective temperature $T$ as in the theory of violent relaxation
based on the Vlasov equation \cite{lb} (see \cite{moczSV} for
the Schr\"odinger-Vlasov
correspondance). In that case, the density decreases at large
distances as $\rho\propto r^{-2}$ \cite{bt}, instead of $r^{-3}$ for the NFW and
Burkert  profiles,
leading
exactly to flat rotation curves for $r\rightarrow +\infty$. For
sufficiently large DM halos, the halo mass-radius relation is
given by \cite{modeldm}
\begin{eqnarray}
\label{intro1}
M_h=1.76\, \Sigma_0 r_h^2,
\end{eqnarray} 
where
\begin{eqnarray}
\label{intro2}
\Sigma_0=\rho_0 r_h=141\, M_{\odot}/{\rm
pc}^2
\end{eqnarray} 
is the universal surface density of DM halos deduced from the observations
\cite{kormendy,spano,donato}. Ultracompact halos like dSphs ($r_h\sim
1\, {\rm kpc}$ and $M_h\sim 10^{8}\, M_\odot$) are dominated by
the quantum core and have almost no atmosphere. They correspond to the
ground state of the quantum model. Large halos like the Medium
Spiral ($r_h\sim
10\, {\rm kpc}$ and $M_h\sim 10^{11}\, M_\odot$) are dominated by the isothermal
atmosphere.

A fundamental problem in the physics of quantum DM halos is to determine
the relation $M_c(M_v)$ between the quantum core mass $M_c$ and the halo mass
$M_v$. In
the case of noninteracting bosons, the scaling $M_c\propto
M_v^{1/3}$  was
obtained numerically by Schive {\it et al.}  \cite{ch3} (see also
\cite{veltmaat}) and explained with a heuristic argument based on a wavelike
``uncertainty principle''. It was then observed by several authors
 \cite{mocz,bbbs,modeldm} that this
relation
could be obtained from a ``velocity dispersion tracing'' relation
according to which
the velocity dispersion in the core $v_c^2\sim GM_c/R_c$ is of the same order as
the velocity
dispersion in the halo $v_v^2\sim GM_v/r_v$. In recent
papers \cite{modeldm,mcmh}, we managed to justify this relation from an
effective
thermodynamic
approach. We considered DM halos with a quantum core and an
homogeneous isothermal
atmosphere in a box of radius $r_h$. We
analytically computed the free energy $F(M_c)$
and the entropy $S(M_c)$ of this core-halo configuration as a function of the
core mass $M_c$. The equilibrium core mass was then determined by extremizing
the free energy 
$F(M_c)$ at fixed halo mass $M_h$, or by extremizing the entropy  $S(M_c)$ at
fixed halo mass $M_h$ and energy $E_h$ (these extremization problems are
equivalent). In this manner, we could obtain the core mass $M_c$ as a function
of the halo mass $M_h$. We showed that the resulting relation is equivalent to
the
``velocity dispersion tracing'' relation $GM_c/R_c\sim GM_v/r_v$. We
could therefore
provide  a
justification of the ``velocity dispersion tracing''  relation from
thermodynamical arguments. We also showed
that the core-halo configuration is a maximum of free energy (instead of being a
minimum)
so that it is unstable in the canonical ensemble. In particular, it has a
negative specific heat which is forbidden in the canonical ensemble. However,
the statistical ensembles are inequivalent for systems with long-range
interactions like gravitational systems \cite{paddy,found,ijmpb}. We could then
show that, if the halo mass is not too large, the
core-halo configuration is a maximum of entropy at fixed mass and
energy so that it
is stable in the microcanonical ensemble. This makes the core-halo
configuration extremely important since it corresponds to the ``most
probable'' configuration of the system (in a thermodynamical sense). Finally,
using the ``velocity dispersion
tracing'' relation, we could generalize the
core mass -- halo mass relation obtained by Schive {\it et al.} 
\cite{ch3} to the
case of DM halos made of bosons with attractive or repulsive
interaction and  to the
case of DM halos made of fermions \cite{modeldm,mcmh}.

Based on these results, we have developed the
following scenario 
which is valid for bosons and
fermions (see Fig. 49 of  \cite{modeldm} for an illustration of this scenario in
the case of bosons with a repulsive self-interaction). There exists a ``minimum
halo'' of mass $(M_h)_{\rm min}$ corresponding to the
ground state of the quantum gas ($T=0$) at which the DM halo is a purely
quantum
object without isothermal atmosphere ($M_c\simeq M_h$). Observations reveal
that $(M_h)_{\rm min}\sim 10^8\, M_{\odot}$. Above a canonical
critical
point $(M_h)_{\rm CCP}$, the DM halos have a core-halo structure with
a quantum core and an isothermal atmosphere. The quantum core may mimic a
galactic nucleus or a large bulge, but it cannot mimic a central black hole
(BH). At
the beginning of this branch, the core-halo structure is stable in the
microcanonical ensemble and the core mass $M_c$ increases with the halo mass
$M_h$.\footnote{We also found a branch along which the
core mass $M_c$ decreases as the halo mass $M_h$ increases. Rapidly, the core
mass becomes negligible and the halos behave as purely isothermal halos without
quantum core.} Above a
microcanonical critical point $(M_h)_{\rm MCP}$, the core-halo structure
becomes thermodynamically unstable and the quantum core is replaced by a
supermassive black hole (SMBH) resulting from a gravothermal catastrophe
\cite{lbw} followed by a
dynamical instability of general relativistic origin \cite{balberg}. In that
process the halo is left undisturbed and conserves its approximately isothermal
structure. The gravothermal catastrophe is a slow (secular) process but it may
be relevant in galactic nuclei or if the DM particle has a
self-interaction \cite{balberg}. In conclusion, we predicted in
\cite{modeldm,mcmh} that DM halos with a mass $(M_h)_{\rm min}<M_h<(M_h)_{\rm
CCP}$ are purely quantum objects  while 
DM halos with a mass $M_h>(M_h)_{\rm CCP}$ have a core-halo structure with an 
approximately isothermal atmosphere. In the latter case, DM
halos with a mass $(M_h)_{\rm CCP}<M_h<(M_h)_{\rm MCP}$ harbour a quantum
core (bosonic soliton or fermion ball) while  DM halos with a mass
$M_h>(M_h)_{\rm MCP}$ harbour a SMBH.
These results are connected to the fundamental existence of a canonical
and a microcanonical critical point in the statistical mechanics of
self-gravitating
systems \cite{ijmpb}.

The previous scenario, if correct, could explain the formation of SMBHs
at the centers of DM halos resulting from the collapse of the quantum core above
a maximum halo mass $(M_h)_{\rm MCP}$. However, it is also possible that DM
halos of any size harbour a SMBH that was present prior to the formation of the
quantum core. Actually, it is observed that most galaxies harbour a SMBH and
that the SMBH mass
$M_{\rm BH}$ is correlated with the velocity
dispersion $\sigma$ of the hot-galaxy bulge, leading to the so-called $M_{\rm
BH}-\sigma$ relation. As a result, the SMBH mass is also correlated  with global
halo properties such as the total halo mass $M_v$, leading to
the BH mass -- halo mass relation $M_{\rm BH}(M_v)$. In that case, we
must revise our scenario \cite{modeldm,mcmh} by taking into account the
influence of the BH on the mass-radius relation of the quantum
core. To
simplify the problem, we shall treat the BH as a point mass. In that case, the
mass-radius relation $M_c(R_c)$ of the quantum core has been obtained in
\cite{epjpbh} by using approximate analytical methods based on a Gaussian ansatz
for the density profile. We shall also use the ``velocity dispersion tracing''
relation $GM_c/R_c\sim GM_h/r_h$, assuming that this relation remains valid in
the presence
of a SMBH. From the BH
mass -- halo mass relation $M_{\rm BH}(M_h)$, the core mass-radius relation
$M_c(R_c)$, the ``velocity dispersion tracing''
relation $GM_c/R_c\sim GM_h/r_h$ and the halo mass-radius relation
$M_h=1.76\, \Sigma_0 r_h^2$, we can obtain the core mass -- halo mass
relation $M_c(M_h)$. This is the subject of the present paper.\footnote{While
our paper was in preparation, we came accross the interesting paper of
Davies and Mocz \cite{dm} who consider fuzzy DM soliton cores (made of
noninteracting bosons) around
SMBHs. These authors assume that the $M_c(M_h)$ relation is unchanged
by the presence of the SMBH. This is valid if the soliton forms early in
the history of the Universe, before the SMBH. In our study, we make the
opposite assumption and study how the $M_c(M_h)$ relation is affected
by the presence of the SMBH. A comparison between the two studies is made in
Sec. \ref{sec_com}. The effects of a SMBH on BECDM
halos have
also been studied  recently by Avilez {\it et al.} \cite{matosbh}, Eby {\it et
al.} \cite{ebybh}, Bar {\it et al.}
\cite{bblp}, Chavanis \cite{epjpbh} and Brax {\it et al.} \cite{braxbh}. On the
other hand,
Desjacques and Nusser \cite{dn} consider the possibility that soliton cores on
galactic
nuclei may mimic SMBHs.} After recalling general results in Sec. \ref{sec_gr},
we treat the
case of bosonic DM halos in Sec. \ref{sec_bdm} and the case of fermionic DM
halos  in Sec. \ref{sec_fdm}.

\section{General results}
\label{sec_gr}

\subsection{The BH mass -- halo mass relation}
\label{bhmhm}

Most galaxies are known to harbour a SMBH of millions of solar masses at
their center. It is found from observations that the BH mass -- halo mass
relation is given by \cite{ferrarese,bandara}
\begin{equation}
\label{bhmhm1}
\frac{M_{\rm BH}}{M_{\odot}}=1.07\times 10^{-12}\left
(\frac{M_v}{M_{\odot}}\right )^{1.55},
\end{equation}
where $M_v$ is the virial halo mass. In our previous papers \cite{modeldm,mcmh},
we have worked with
the halo mass $M_h$ defined such that $M_h$ is the mass contained within the
sphere of radius $r_h$ where the central density has been divided by four
(i.e. $\rho_h=\rho_0/4$). These two masses are related by \cite{mcmh}
\begin{eqnarray}
\label{bhmhm2}
\frac{M_h}{(M_h)_{\rm min}}\sim B \left\lbrack
\frac{M_v}{(M_h)_{\rm
min}}\right\rbrack^{4/3},
\end{eqnarray}
where
\begin{eqnarray}
\label{bhmhm3}
B=\frac{1}{1.76\, \Sigma_0}\left\lbrack \frac{4}{3}\pi
\zeta(0)\rho_{m,0}\right\rbrack^{2/3} (M_h)_{\rm min}^{1/3}.
\end{eqnarray}
In this expression, $\rho_{m,0}=2.66\times 10^{-24}\, {\rm
g \, m^{-3}}$ is the present background matter density in the Universe,
$\zeta(0)$ is a prefactor of order $350$, and $(M_h)_{\rm min}$ is
the minimum mass of the DM halos observed in the Universe. If we take
$(M_h)_{\rm min}=10^8\, M_{\odot}$, a choice that we will make subsequently, we
obtain $B=2.79\times
10^{-3}$. Therefore
\begin{eqnarray}
\label{bhmhm2b}
\frac{M_h}{(M_h)_{\rm min}}\sim  2.79\times
10^{-3}\left\lbrack
\frac{M_v}{(M_h)_{\rm
min}}\right\rbrack^{4/3}.
\end{eqnarray}
Combining the foregoing relations, we get
\begin{equation}
\label{bhmhm4}
\frac{M_{\rm BH}}{(M_h)_{\rm min}}=2.69\times 10^{-8}\left
(\frac{M_v}{(M_h)_{\rm min}}\right )^{1.55}
\end{equation}
and
\begin{equation}
\label{bhmhm5}
\frac{M_{\rm BH}}{(M_h)_{\rm min}}=2.51\times 10^{-5}\left
(\frac{M_h}{(M_h)_{\rm min}}\right )^{1.16}.
\end{equation}
For the sake of generality, we write the BH mass -- halo mass relation under
the form
\begin{equation}
\label{bhmhm6}
\frac{M_{\rm BH}}{(M_h)_{\rm min}}=A\left
(\frac{M_h}{(M_h)_{\rm min}}\right )^{a},
\end{equation}
where the constants $A$ and $a$ can be updated if necessary. From the results
of \cite{bandara}, their values are $A=2.51\times 10^{-5}$ and $a=1.16$.

\subsection{Velocity dispersion tracing relation }
\label{vdtr}

To obtain the core mass --  halo mass relation $M_c(M_h)$ we shall use the
velocity dispersion tracing relation 
\begin{eqnarray}
\label{mcmh1}
v_c^2\sim v_h^2 \quad {\rm or} \quad M_c\sim \frac{R_c}{r_h}M_h,
\end{eqnarray}
stating that the
velocity dispersion in the core $v_c^2\sim GM_c/R_c$ is of the same order as the
velocity dispersion in the halo $v_h^2\sim GM_h/r_h$. This relation was
introduced in \cite{mocz,bbbs,modeldm}. It was shown to
reproduce the core mass --  halo mass relation obtained numerically by
Schive {\it et al.} \cite{ch3} for noninteracting bosons. In recent papers
\cite{modeldm,mcmh}, we have provided a
justification of this relation from an effective thermodynamic approach by
determining the core mass that maximizes the entropy of the DM halo at
fixed total mass and energy. We showed that this relation remains valid for
self-interacting bosons and for fermions. In the
present paper, we assume that it also remains valid when the halo contains a
central BH. Using the halo mass -- 
radius relation $M_h=1.76\, \Sigma_0
r_h^2$ [see Eq. (\ref{intro1})],\footnote{This relation
presents the
fundamental scaling $M_h\propto r_h^2$ reflecting the universality of the
surface density $\Sigma_0\sim M_h/r_h^2$ of DM halos [see Eq.
(\ref{intro2})].}  we can rewrite Eq. (\ref{mcmh1}) as
\begin{equation}
\label{mcmh2}
\frac{M_c}{R_c}\sim\sqrt{1.76\, \Sigma_0 M_h}.
\end{equation}

\section{Bosonic DM halos}
\label{sec_bdm}

In this section, we obtain the core mass -- halo mass relation of bosonic DM
halos in the presence of a central BH by combining the velocity
dispersion tracing relation (\ref{mcmh2}) with the core mass-radius
relation of a self-gravitating BEC at $T=0$.

\subsection{Core mass-radius relation}
\label{sec_mrr}

A self-gravitating BEC is basically described by the
Gross-Pitaevskii-Poisson (GPP) equations (see, e.g., \cite{prd1}). Far from the
Schwarzschild radius, we can treat the central BH as a point mass creating an
external
potential $\Phi_{\rm ext}=-GM_{\rm BH}/r$. Using a Gaussian ansatz for the wave
function, we found in \cite{epjpbh} that the approximate
mass-radius relation of a self-gravitating BEC at $T=0$ (ground state) with a
central BH is given
by
\begin{equation}
\label{mrr1}
M_c=\frac{2\sigma}{\nu}\frac{\frac{\hbar^2}{Gm^2R_c}-\frac{\lambda}{2\sigma}M_ {
\rm BH}}{1-\frac{6\pi\zeta
a_s\hbar^2}{\nu Gm^3 R_c^2}}
\end{equation}
with the coefficients $\sigma=3/4$, $\zeta=1/(2\pi)^{3/2}$,
$\nu=1/\sqrt{2\pi}$ and $\lambda=2/\sqrt{\pi}$. 
Inversely, the radius of the BEC can be expressed in terms of its
mass by
\begin{eqnarray}
\label{mrr2}
R_c&=&\frac{\sigma}{\nu}\frac{\hbar^2}{Gm^2}\frac{1}{M_c+\frac{\lambda}{\nu}
M_{\rm BH}}\nonumber\\
&\times& \left
\lbrack 1\pm\sqrt{1+\frac{6\pi\zeta\nu}{\sigma^2}\frac{Gm
a_s}{\hbar^2}M_c\left (M_c+\frac{\lambda}{\nu}
M_{\rm BH}
\right )} \right \rbrack\quad\quad
\end{eqnarray}
with $+$ when $a_s>0$ and with $\pm$ when $a_s<0$. When $M_{\rm BH}=0$, we
recover the 
approximate mass-radius relation obtained in \cite{prd1}. The results  of
\cite{epjpbh,prd1} apply to the  ``minimum halo''  which has no isothermal
atmosphere (ground
state) and to the
quantum core of larger DM halos which have an isothermal atmosphere.
For a given value of the core mass $M_c$, the
effect of the BH
is to decrease core radius $R_c$ \cite{epjpbh}.

For
noninteracting BECs
($a_s=0$), the mass-radius relation (\ref{mrr1}) reduces to
\begin{equation}
\label{mrr3}
M_c=\frac{2\sigma}{\nu}\left
(\frac{\hbar^2}{Gm^2R_c}-\frac{\lambda}{2\sigma } M_ {
\rm BH}\right )
\end{equation}
or, inversely,
\begin{eqnarray}
\label{mrr4}
R_c=\frac{2\sigma}{\nu}\frac{\hbar^2}{Gm^2}\frac{1}{M_c+\frac{\lambda}{\nu}
M_{\rm BH}}.
\end{eqnarray}
The radius $R_c$ decreases as the mass $M_c$ increases, going from the
gravitational Bohr radius\footnote{When the gravitational
attraction of the BH dominates the self-gravity of the quantum core,
the GPP equations reduce (in the noninteracting case) to the Schr\"odinger
equation for the hydrogen atom \cite{epjpbh}.}
\begin{eqnarray}
\label{mrr5}
R_{\rm
B}=\frac{2\sigma}{\lambda}\frac{\hbar^2}{GM_{\rm
BH}m^2}
\end{eqnarray}
when $M_c\rightarrow 0$ down to zero when $M_c\rightarrow +\infty$ (see Fig.
3 of
\cite{epjpbh}). All these configurations are stable. 
For a given value of the core mass $M_c$, the core radius decreases as
the BH mass increases, going from  
\begin{eqnarray}
\label{mrr5q}
R_c=\frac{2\sigma}{\nu}\frac{\hbar^2}{Gm^2M_c}
\end{eqnarray}
when $M_{\rm BH}=0$ to $R_c=R_{\rm B}\propto M_{\rm BH}^{-1}\rightarrow 0$
when the BH dominates ($M_{\rm BH}\rightarrow
+\infty$). In the later case, the core radius is inversely
proportional to the BH mass [see Eq.
(\ref{mrr5})] rather than the core mass [see Eq. (\ref{mrr5q})].

We now consider the
repulsive case ($a_s>0$). In the TF limit ($\hbar=0$), the mass-radius relation
(\ref{mrr1})
reduces to
\begin{equation}
\label{mrr1tf}
M_c=\frac{-\frac{\lambda}{\nu}M_{\rm BH}}{1-\frac{6\pi\zeta
a_s\hbar^2}{\nu Gm^3 R_c^2}}
\end{equation}
or, inversely,
\begin{eqnarray}
\label{mrr4tf}
R_c=\left
(\frac{6\pi\zeta}{\nu}\right
)^{1/2}\left
(\frac{a_s\hbar^2}{Gm^3}\right
)^{1/2}\frac{1}{\sqrt{1+\frac{\lambda}{\nu}\frac{M_{\rm BH}}{M_c}}}.
\end{eqnarray}
The radius $R_c$ increases as the mass $M_c$ increases, going from $R_c=0$ when
$M_c=0$ to the TF radius
\begin{equation}
\label{mrr7}
R_{\rm TF}=\left (\frac{6\pi\zeta}{\nu}\right )^{1/2}\left
(\frac{a_s\hbar^2}{Gm^3}\right )^{1/2}
\end{equation}
when $M\rightarrow +\infty$. 
 (see Fig. 4 of
\cite{epjpbh}). All these configurations are stable. 
For a given value of the core mass $M_c$, the core radius decreases as
the BH mass increases, going from  $R=R_{\rm TF}$ when $M_{\rm BH}=0$ to  
\begin{eqnarray}
\label{mrr7q}
R_c= \left
(\frac{6\pi\zeta}{\lambda}\right
)^{1/2}\left
(\frac{a_s\hbar^2 M_c}{Gm^3 M_{\rm BH}}\right
)^{1/2}\rightarrow 0
\end{eqnarray}
when the BH dominates ($M_{\rm BH}\rightarrow +\infty$). We now consider the 
general case of a
repulsive self-interaction. We must consider two cases. When $M_{\rm
BH}<(M_{\rm
BH})_{*}$ with
\begin{eqnarray}
\label{mrr6}
(M_{\rm BH})_{*}=\frac{2\nu}{\lambda}\left (\frac{\sigma^2}{6\pi\zeta\nu}\right
)^{1/2}\frac{\hbar}{\sqrt{Gm|a_s|}},
\end{eqnarray}
the radius $R_c$ decreases as the mass $M_c$ increases, going from the
gravitational Bohr radius $R_{\rm B}$ when $M_c\rightarrow 0$ down to the TF
radius $R_{\rm TF}$
when $M_c\rightarrow +\infty$ (see Fig. 5 of
\cite{epjpbh}).  When $M_{\rm BH}>(M_{\rm
BH})_{*}$, the radius $R_c$ increases as the mass
$M_c$ increases, going from the
gravitational Bohr radius $R_{\rm B}$ when $M\rightarrow 0$ up to the TF radius
$R_{\rm TF}$ when $M\rightarrow +\infty$ (see Fig. 6 of
\cite{epjpbh}).  All these configurations are stable. 

In the
attractive case
($a_s<0$)  the mass-radius relation is nonmonotonic
(see Fig. 10 of
\cite{epjpbh}).  There is a maximum mass $(M_c)_{\rm max}(M_{\rm BH})$ given by
\begin{eqnarray}
\label{mrr8}
\frac{(M_c)_{\rm max}(M_{\rm BH})}{(M_{c})_{\rm
max}}&=&-\frac{\lambda}{2\nu}\frac{M_{\rm BH}}{(M_{c})_{\rm
max}}\nonumber\\
&+&\sqrt{1+\frac{\lambda^2}{4\nu^2}\left (\frac{M_{\rm BH}}{(M_{c})_{\rm
max}}\right )^2},
\end{eqnarray}
where
\begin{equation}
\label{mrr9}
(M_{c})_{\rm max}=\left (\frac{\sigma^2}{6\pi\zeta\nu}\right
)^{1/2}\frac{\hbar}{\sqrt{Gm|a_s|}}
\end{equation}
is the maximum mass of a self-gravitating BEC with an attractive
self-interaction without a central BH \cite{prd1}. The maximum mass decreases
as the mass of the BH increases (see Fig. 11 of \cite{epjpbh}). The radius
$(R_c)_{*}(M_{\rm BH})$ corresponding
to $(M_c)_{\rm max}(M_{\rm BH})$ is given by
\begin{equation}
\label{mrr10}
\frac{(R_c)_{*}(M_{\rm BH})}{(R_c)_*}=\frac{(M_c)_{\rm max}(M_{\rm
BH})}{(M_{c})_{\rm
max}},
\end{equation}
where
\begin{equation}
\label{mrr11}
(R_c)_*=\left (\frac{6\pi\zeta}{\nu}\right )^{1/2}\left
(\frac{|a_s|\hbar^2}{Gm^3}\right )^{1/2}
\end{equation}
is the radius corresponding to $(M_c)_{\rm max}$ in the absence of a central BH
\cite{prd1}. No equilibrium state exist with a mass $M_c>(M_{c})_{\rm
max}(M_{\rm BH})$. For
$M_c<(M_{c})_{\rm max}(M_{\rm BH})$ the branch $R_c>(R_c)_*(M_{\rm BH})$
(corresponding to the solutions
(\ref{mrr2}) with the sign $+$) is stable
while the branch $R_c<(R_c)_*(M_{\rm BH})$ (corresponding to the solutions
(\ref{mrr2}) with
the sign $-$) is 
unstable. On the stable branch  the radius $R_c$ decreases as the mass
$M_c$ increases, going from the
gravitational Bohr radius $R_{\rm B}$ when $M_c\rightarrow 0$ down to the
minimum
stable radius $(R_c)_{*}(M_{\rm BH})$ when $M_c\rightarrow (M_c)_{\rm
max}(M_{\rm
BH})$.

{\it Remark:} When $a_s\ge 0$ all the configurations, with any
value of $M$, are stable. Although the central BH enhances the
gravitational attraction and reduces the radius of the BEC, it
does not destabilize the system. When
$a_s<0$, only the configurations below the maximum mass $(M_c)_{\rm
max}(M_{\rm BH})$ and above the minimum radius $(R_c)_{*}(M_{\rm BH})$ are
stable
in continuity with the case without central BH \cite{prd1}.

\subsection{The core mass -- halo mass relation}
\label{sec_mcmh}

Substituting the core mass -- radius relation (\ref{mrr1}) into Eq.
(\ref{mcmh2}), we obtain the
second degree equation
\begin{eqnarray}
\label{mcmh3}
\left (\frac{M_c}{(M_h)_{\rm min,0}}\right
)^2+\frac{\lambda}{\nu}\frac{M_{\rm BH}}{(M_h)_{\rm min,0}}\frac{M_c}{(M_h)_{\rm
min,0}}\nonumber\\
=\left (\frac{M_h}{(M_h)_{\rm min,0}}\right
)^{1/2}\left\lbrack 1+\frac{a_s}{a_*}\left (\frac{M_h}{(M_h)_{\rm min,0}}\right
)^{1/2}\right\rbrack
\end{eqnarray}
determining the core mass $M_c$ as a function of the halo mass $M_h$ in the
presence of a central BH of mass $M_{\rm BH}(M_h)$ given by Eq.
(\ref{bhmhm6}).
Following
our previous paper \cite{mcmh} we have introduced the mass scale
\begin{equation}
\label{mcmh4}
(M_h)_{\rm min,0}=\frac{2^{2/3}\sigma^{2/3}}{\nu^{2/3}\alpha^{1/3}}\left
(\frac{\hbar^4\Sigma_0}{G^2m^4}\right )^{1/3}
\end{equation}
and the scattering length scale
\begin{equation}
\label{mcmh5}
a_*=\frac{2^{2/3}\sigma^{2/3}\alpha^{2/3}\nu^{1/3}}{6\pi\zeta}\left
(\frac{Gm^5}{\hbar^2\Sigma_0^2}\right )^{1/3},
\end{equation}
where $\alpha=1/1.76$. Physically
$(M_h)_{\rm min,0}$ gives the minimum mass (ground
state) of a noninteracting self-gravitating BEC without central BH. On
the other hand, the scattering length $a_*$ determines the transition between
the noninteracting regime and
the TF regime (for a repulsive self-interaction) or  the transition between
the noninteracting regime and the collapse regime 
(for an attractive self-interaction). The
solution of Eq. (\ref{mcmh3}) is
\begin{eqnarray}
\label{mcmh6}
\frac{M_c}{(M_h)_{\rm min,0}}=-\frac{\lambda}{2\nu}\frac{M_{\rm BH}}{(M_h)_{\rm
min,0}}
+\Biggl\lbrace \frac{\lambda^2}{4\nu^2}\left (\frac{M_{\rm BH}}{(M_h)_{\rm
min,0}}\right )^2\nonumber\\
+ \left (\frac{M_h}{(M_h)_{\rm min,0}}\right
)^{1/2}\left\lbrack 1+\frac{a_s}{a_*}\left (\frac{M_h}{(M_h)_{\rm min,0}}\right
)^{1/2}\right\rbrack \Biggr\rbrace^{1/2}.\quad
\end{eqnarray}
In the absence of BH we recover the relation 
\begin{equation}
\label{mcmh7}
\frac{M_c}{(M_h)_{\rm min,0}}=\left (\frac{M_h}{(M_h)_{\rm min,0}}\right
)^{1/4}\sqrt{1+\frac{a_s}{a_*}\left (\frac{M_h}{(M_h)_{\rm min,0}}\right
)^{1/2}}
\end{equation}
obtained in \cite{mcmh}. When the BH dominates, we
get
\begin{eqnarray}
\label{mcmh8}
\frac{M_c}{(M_h)_{\rm
min,0}}
=\frac{\nu}{\lambda}\frac{(M_h)_{\rm min,0}}{M_{\rm
BH}}\left (\frac{M_h}{(M_h)_{\rm min,0}}\right
)^{1/2}\nonumber\\
\times\left\lbrack 1+\frac{a_s}{a_*}\left (\frac{M_h}{(M_h)_{\rm
min,0}}\right
)^{1/2}\right\rbrack.
\end{eqnarray}

\subsection{The minimum halo (ground state)}
\label{sec_mh}

Setting $M_c=M_h$ in Eq. (\ref{mcmh6}) we obtain 
the minimum halo mass
$(M_h)_{\rm
min}$ (ground state) as a function of $a_s/a_*$ in the presence of a central BH.
However, for the minimum halo, $M_{\rm
BH}/(M_h)_{\rm min}\sim 2.51\times 10^{-5}\ll 1$  and,
consequently, we can neglect the effect
of the BH. The effect
of the BH
will be important only for larger halos (see below). Therefore, the results of
Secs. VII.B and VII.C of \cite{mcmh} are unchanged. In particular, we obtain the
following
relation
\begin{equation}
\label{mh1}
\frac{a_s}{a_*}=\frac{(M_{h})_{\rm min}}{(M_{h})_{\rm
min,0}}-\sqrt{\frac{(M_{h})_{\rm min,0}}{(M_{h})_{\rm min}}}
\end{equation}
between  the minimum halo mass
$(M_h)_{\rm
min}$ and the scattering length $a_s/a_*$ of the DM particle. This relation is
plotted in Fig. 11 of \cite{mcmh}. When
$a_s=0$ we
have $(M_h)_{\rm min}=(M_h)_{\rm min,0}$. When $a_s\ge 0$, $(M_h)_{\rm
min}$ is larger than $(M_h)_{\rm min,0}$. When $a_s\le 0$, $(M_h)_{\rm
min}$ is smaller than $(M_h)_{\rm min,0}$. When $a_s\ge 0$, the minimum halo is
always stable. However, when 
$a_s<0$, it is stable only if $M_c<(M_c)_{\rm max}$ where $(M_c)_{\rm max}$ is
given by \cite{mcmh}:
\begin{equation}
\label{mh2}
\frac{(M_{c})_{\rm max}}{(M_h)_{\rm min,0}}=\frac{1}{2}\left
(\frac{a_*}{|a_s|}\right )^{1/2}.
\end{equation}
As a result, the  minimum halo is stable provided that $a_s\ge (a_s)_c$
with
\begin{equation}
\label{mh3}
\frac{(a_s)_c}{a_*}=-\frac{1}{2^{2/3}}.
\end{equation}
For $a_s=(a_s)_c$, the minimum halo is critical. It has a mass
\begin{equation}
\label{mh4}
\frac{(M_{h})_{\rm
min,c}}{(M_h)_{\rm min,0}}=\frac{1}{2^{2/3}}.
\end{equation}
Consequently, for a scattering length $(a_s)_c\le a_s\le 0$, the mass of the
(stable) minimum halo is in the range $(M_{h})_{\rm
min,c}\le M_h\le (M_{h})_{\rm
min,0}$.

In principle, the  minimum halo mass $(M_h)_{\rm min}$ can be obtained from the
observations. Its value is not known precisely but it is of the order of
$10^8\, M_{\odot}$ corresponding to dSphs like Fornax. The 
minimum halo mass
$(M_h)_{\rm min}$ is the only unknown parameter of our theory. In
the
following, to fix the ideas, we will take
\begin{equation}
\label{mh5}
(M_h)_{\rm min}=10^8\, M_{\odot}.
\end{equation}
Our
formula are general but the numerical applications will
slightly change if other, more accurate, values of $(M_h)_{\rm min}$ are used
instead of Eq. (\ref{mh5}). We also recall that our approach is approximate
because it is based on a Gaussian ansatz for the density profile of DM halos.
Again, it could be improved by using the exact core mass-radius
relation.  However, our main aim in this paper is to present the general ideas,
so our approximate analytical approach is sufficient for our purposes.

Once the mass $(M_h)_{\rm min}$ of the minimum halo is fixed, Eq. (\ref{mh1})
determines
the relation betwen
the DM particle mass $m$ and the scattering length $a_s$. This relation can be
written as \cite{mcmh}:
\begin{equation}
\label{mh6}
\frac{a_s}{a'_*}=\left (\frac{m}{m_0}\right )^3-\frac{m}{m_0},
\end{equation}
where we have introduced the particle mass scale
\begin{equation}
\label{mh7}
m_0=\frac{2^{1/2}\sigma^{1/2}}{\nu^{1/2}\alpha^{1/4}}\frac{\hbar\Sigma_0^{1/4}}{
G^ {1/2}(M_h)_{\rm min}^{3/4}}
\end{equation}
and the scattering length scale
\begin{equation}
\label{mh8}
a'_*=\frac{2^{3/2}\sigma^{3/2}\alpha^{1/4}}{\nu^{1/2}6\pi\zeta}\frac{\hbar}{G^{
1/2}\Sigma_0^{1/4}(M_h)_{\rm min}^{5/4}}.
\end{equation}
For $(M_h)_{\rm min}=10^8\, M_{\odot}$ we obtain
\begin{equation}
\label{mh9}
m_0=2.25\times
10^{-22}\, {\rm eV}/c^2
\end{equation}
and
\begin{equation}
\label{mh10}
a'_*=4.95\times 10^{-62}\, {\rm fm}.
\end{equation}
Physically, $m_0$ represents the mass of the DM particle in the noninteracting
case ($a_s=0$) which is consistent with a minimum halo of mass $(M_h)_{\rm
min}$. On the other
hand, the scattering length $a'_*$ determines the transition between
the noninteracting regime and
the TF regime (for a repulsive self-interaction) or  the transition between
the noninteracting regime and the collapse regime 
(for an attractive self-interaction). According to Eq. (\ref{mh6}), our results
depend only on the scattering length $a_s$ of the DM particle: its mass $m$  is
then  automatically determined by Eq. (\ref{mh6}).\footnote{Of course, we could
take
the
opposite viewpoint and consider that the scattering length $a_s$ is determined
by the mass $m$ so as to be  consistent with a minimum halo of mass $(M_h)_{\rm
min}$. However, it is more convenient to take $a_s$ as the control parameter.} 

The relation (\ref{mh6}) is plotted in Fig. 13 of \cite{mcmh}. When $a_s=0$ we
have $m=m_0$. When $a_s\ge 0$, $m$ is larger than $m_0$.
When $a_s\le 0$, $m$ is smaller than $m_0$. When $a_s\le
0$, the minimum halo is stable only if $a_s>(a_s)_c$ with
\begin{equation}
\label{mh11}
\frac{(a_s)_{c}}{a'_*}=-\frac{1}{2^{3/2}}.
\end{equation}
For $a_s=(a_s)_c$, the minimum halo is critical. This corresponds to a particle
mass
\begin{equation}
\label{mh12}
\frac{m_c}{m_0}=\frac{1}{\sqrt{2}}.
\end{equation}
Consequently, for a scattering length $(a_s)_c\le a_s\le 0$, the mass of the DM
particle is in the range $m_c\le m\le m_0$.

{\it Remark:} For bosons with an attractive self-interaction, like the axion
\cite{marshrevue}, it is more
convenient to express the results in terms of the decay constant
(see, e.g., \cite{phi6})
\begin{equation}
\label{mh13}
f=\left (\frac{\hbar c^3 m}{32\pi |a_s|}\right )^{1/2},
\end{equation}
rather than the scattering length $a_s$. In that case, the relation (\ref{mh6})
can be
rewritten as
\begin{equation}
\label{mh14}
\frac{f}{f'_*}=\frac{1}{\sqrt{1-\left (\frac{m}{m_0}\right )^2}},
\end{equation}
where we have introduced the energy scale \cite{mcmh}:
\begin{equation}
\label{mh14b}
f'_*=\frac{(6\pi\zeta)^{1/2}}{8\pi^{1/2}\sigma^{1/2}\alpha^{1/4}}\hbar^{1/2}
\Sigma_0^{1/4}(M_h)_{\rm min}^{1/4}c^{3/2}.
\end{equation}
For $(M_h)_{\rm min}=10^8\, M_{\odot}$ we obtain
\begin{equation}
\label{mh15}
f'_*=9.45\times 10^{13}\, {\rm GeV}.
\end{equation}
The relation (\ref{mh14}) is plotted in Fig. 14 of \cite{mcmh}. The minimum halo
is stable only if $f>f_c$ with
\begin{equation}
\label{mh11b}
\frac{f_{c}}{f'_*}=\sqrt{2}.
\end{equation}
For $f=f_c$, the minimum halo is critical. This corresponds to a particle
mass $m_c$. When $f\rightarrow +\infty$ we
have $m\rightarrow m_0$.

\subsection{Procedure to determine the core mass -- halo mass relation}
\label{sec_proc}

In order to determine the core mass -- halo mass relation for a given value of 
the scattering length $a_s$, we need the following relation
\cite{mcmh}
\begin{equation}
\label{proc1}
\frac{a_s}{a_*}=\left (\frac{m}{m_0}\right)^{4/3}-\left
(\frac{m_0}{m}\right)^{2/3},
\end{equation}
which is obtained from Eq. (\ref{mh6}) by normalizing the  scattering length
$a_s$ by the  scattering length scale $a_*$ introduced in Sec. \ref{sec_mcmh}
(see \cite{mcmh} for details).
We can now proceed in the following manner. For a given value of $a_s/a'_*$ we
can obtain $m/m_0$ from Eq. (\ref{mh6}). Then, we get $a_s/a_*$ from Eq.
(\ref{proc1}) and    
$(M_h)_{\rm min}/(M_h)_{\rm min,0}$ from Eq. (\ref{mh1}). We can then plot
$M_c/(M_h)_{\rm min}$ as a function of
$M_{h}/(M_h)_{\rm min}$  by using Eqs. (\ref{bhmhm6}) and (\ref{mcmh6}). We
stress that this
procedure yields a ``universal'' curve $M_c/(M_h)_{\rm min}$ vs
$M_{h}/(M_h)_{\rm min}$ for a given value of $a_s/a'_*$. The only unknown
parameter of our model is the mass
$(M_h)_{\rm min}$ of the minimum halo which determines the scales
$m_0$ and $a'_*$ according to Eqs. (\ref{mh7}) and (\ref{mh8}).\footnote{We
have also assumed that the surface density of the DM halos
is universal [see Eq. (\ref{intro2})]. If this were not the case, our general
model would
remain valid but the problem would depend on two parameters, $M_h$ and $r_h$,
instead of just
$M_h$.} The minimum halo mass is not known with precision but it is
of order $(M_h)_{\rm min}\sim 10^8\, M_{\odot}$.  In the numerical applications,
we will assume that $(M_h)_{\rm min}= 10^8\, M_{\odot}$ [see
Eq. (\ref{mh5})]. If we change the value of the minimum halo mass, the
normalized curve $M_c/(M_h)_{\rm min}(M_h/(M_h)_{\rm min})$ for a given value of
$a_s/a'_*$ remains the same: only the scales change.

{\it Remark:} For convenience, we have proceeded the other way
round. We have fixed a value of $(M_h)_{\rm min}/(M_h)_{\rm min,0}$, determined 
$a_s/a_*$ from Eq. (\ref{mh1}), and plotted $M_c/(M_h)_{\rm min}$ as a
function of
$M_{h}/(M_h)_{\rm min}$   by using Eqs. (\ref{bhmhm6}) and (\ref{mcmh6}). We
have
then used Eqs. (\ref{proc1}) and (\ref{mh6}) to obtain the
values of $m/m_0$ and  $a_s/a'_*$ corresponding to our choice of $(M_h)_{\rm
min}/(M_h)_{\rm min,0}$.

\subsection{Noninteracting bosons}
\label{sec_ni}

\begin{figure}[!h]
\begin{center}
\includegraphics[clip,scale=0.3]{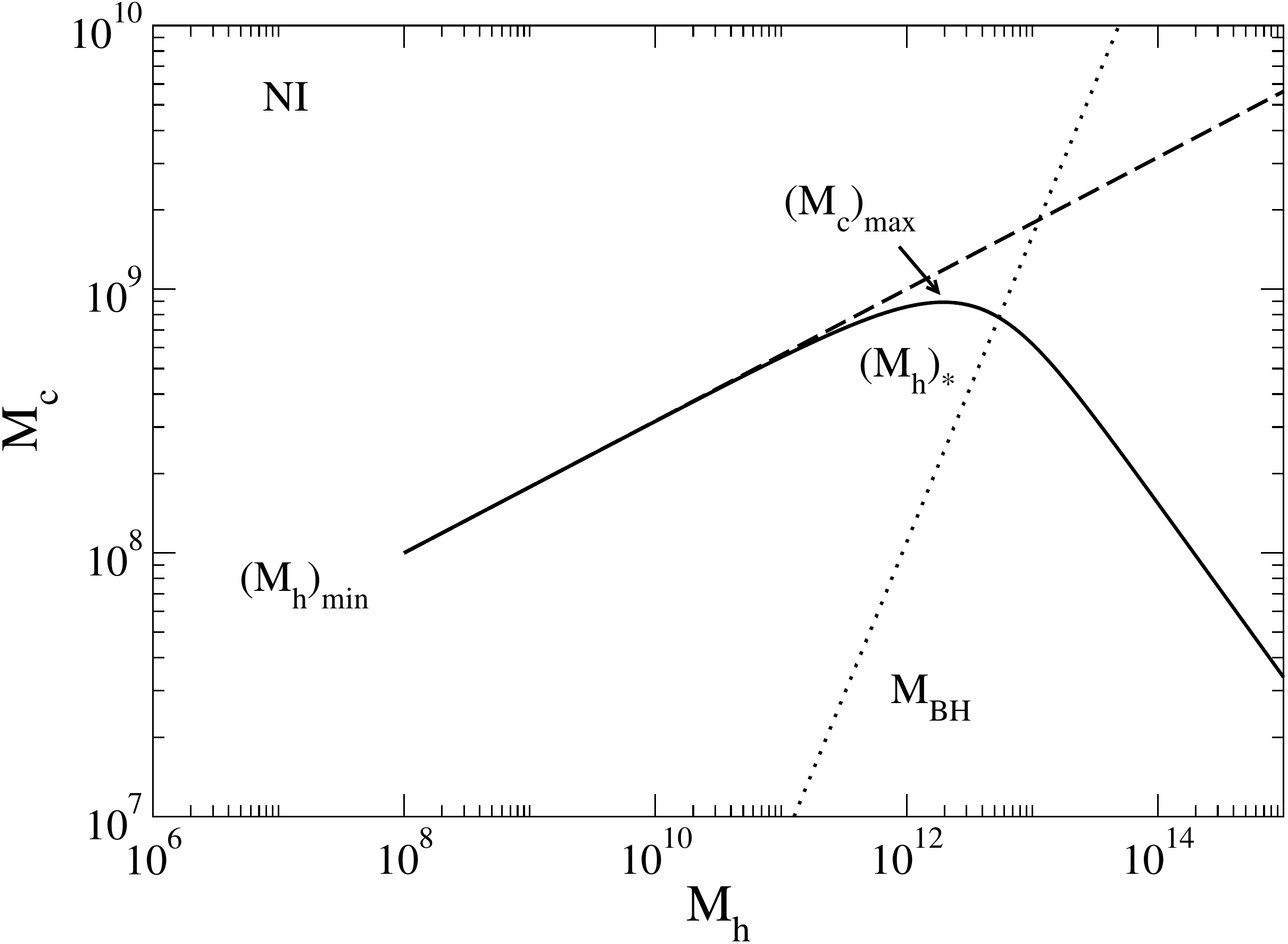}
\caption{Core mass $M_c$ as a function of the halo mass $M_h$ (solid line) for
noninteracting bosons ($a_s=0$; $m=m_0=2.25\times
10^{-22}\, {\rm eV}/c^2$). The mass is
normalized by $M_{\odot}$.  The function $M_c(M_h)$ presents a
maximum core mass $(M_c)_{\rm max,0}=8.92\times 10^8\,
M_{\odot}$ at $(M_h)_{*,0}=1.96\times 10^{12}\, M_{\odot}$. All the
configurations are
stable. We have also represented the
relation $M_c(M_h)$ without BH (dashed line) and the
relation $M_{\rm BH}(M_h)$ (dotted line).}
\label{mhmcNI}
\end{center}
\end{figure}

For noninteracting bosons ($a_s=0$), the core mass -- halo
mass
relation (\ref{mcmh6}) reduces to 
\begin{eqnarray}
\label{ni1}
&&\frac{M_c}{(M_h)_{\rm min}}=-\frac{\lambda}{2\nu}\frac{M_{\rm
BH}}{(M_h)_{\rm
min}}\nonumber\\
&+&\sqrt{\frac{\lambda^2}{4\nu^2}\left (\frac{M_{\rm BH}}{(M_h)_{\rm
min}}\right )^2
+ \left (\frac{M_h}{(M_h)_{\rm min}}\right
)^{1/2}},
\end{eqnarray}
where $M_{\rm BH}$ is given as a function of $M_h$  by Eq. (\ref{bhmhm6}).
More generally, this relation is valid for $|a_s|\ll
a'_*$. It is plotted in Fig. \ref{mhmcNI}. There is a maximum
core mass
\begin{eqnarray}
\label{ni2}
(M_c)_{\rm max,0}=8.92\times 10^8\, M_{\odot},
\end{eqnarray}
corresponding to a halo mass
\begin{eqnarray}
\label{ni3}
(M_h)_{*,0}=1.96\times 10^{12}\, M_{\odot}
\end{eqnarray}
and a BH mass
\begin{eqnarray}
\label{ni4}
(M_{\rm BH})_{*,0}=2.39\times 10^8\, M_{\odot}.
\end{eqnarray}
The effect of the BH becomes
important when $M_{\rm BH}/(M_h)_{\rm min}\sim ({M_h}/(M_h)_{\rm min})^{1/4}$,
i.e. $M_{\rm BH}\sim M_c$, corresponding  to $(M_h)_{*,0}\sim 10^{12}\,
M_{\odot}$. When $M_h\ll (M_h)_{*,0}$ the effect of the BH is
negligible and we recover the scaling 
\begin{eqnarray}
\label{ni5}
\frac{M_c}{(M_h)_{\rm min}}=\left
(\frac{M_h}{(M_h)_{\rm min}}\right
)^{1/4}
\end{eqnarray}
obtained in \cite{mcmh}. When $M_h\gg (M_h)_{*,0}$ the BH
dominates 
and we get
the
scaling
\begin{eqnarray}
\label{ni6}
\frac{M_c}{(M_h)_{\rm min}}&\sim& \frac{\nu}{\lambda}\left
(\frac{M_h}{(M_h)_{\rm min}}\right )^{1/2}\frac{(M_h)_{\rm min}}{M_{\rm
BH}}\nonumber\\
&\sim& \frac{\nu}{\lambda A}\left
(\frac{(M_h)_{\rm min}}{M_h}\right )^{a-1/2}.
\end{eqnarray}
This relation can be directly obtained from Eqs. (\ref{mcmh2})
and (\ref{mrr5}). It exhibits a critical index $a_{0}=1/2$. When
$a>a_{0}$ the core
mass
decreases with $M_h$ and when  $a<a_{0}$ the core mass
increases with $M_h$. For the measured value $a=1.16$, we are in the first case.
For a DM halo of mass $M_h=10^{12}\,
M_{\odot}$ similar to the one that surrounds our Galaxy, we obtain a core
mass $M_c=8.55\times 10^{8}\,
M_{\odot}$ a little smaller than the value $M_c=10^{9}\,
M_{\odot}$  obtained in \cite{mcmh}  in the
absence of a central BH (we have taken $(M_h)_{\rm min}=10^{8}\,
M_{\odot}$).

\subsection{Bosons with a repulsive self-interaction}
\label{sec_tf}

In this section, we consider the case of bosons with a repulsive
self-interaction ($a_s>0$).

\begin{figure}[!h]
\begin{center}
\includegraphics[clip,scale=0.3]{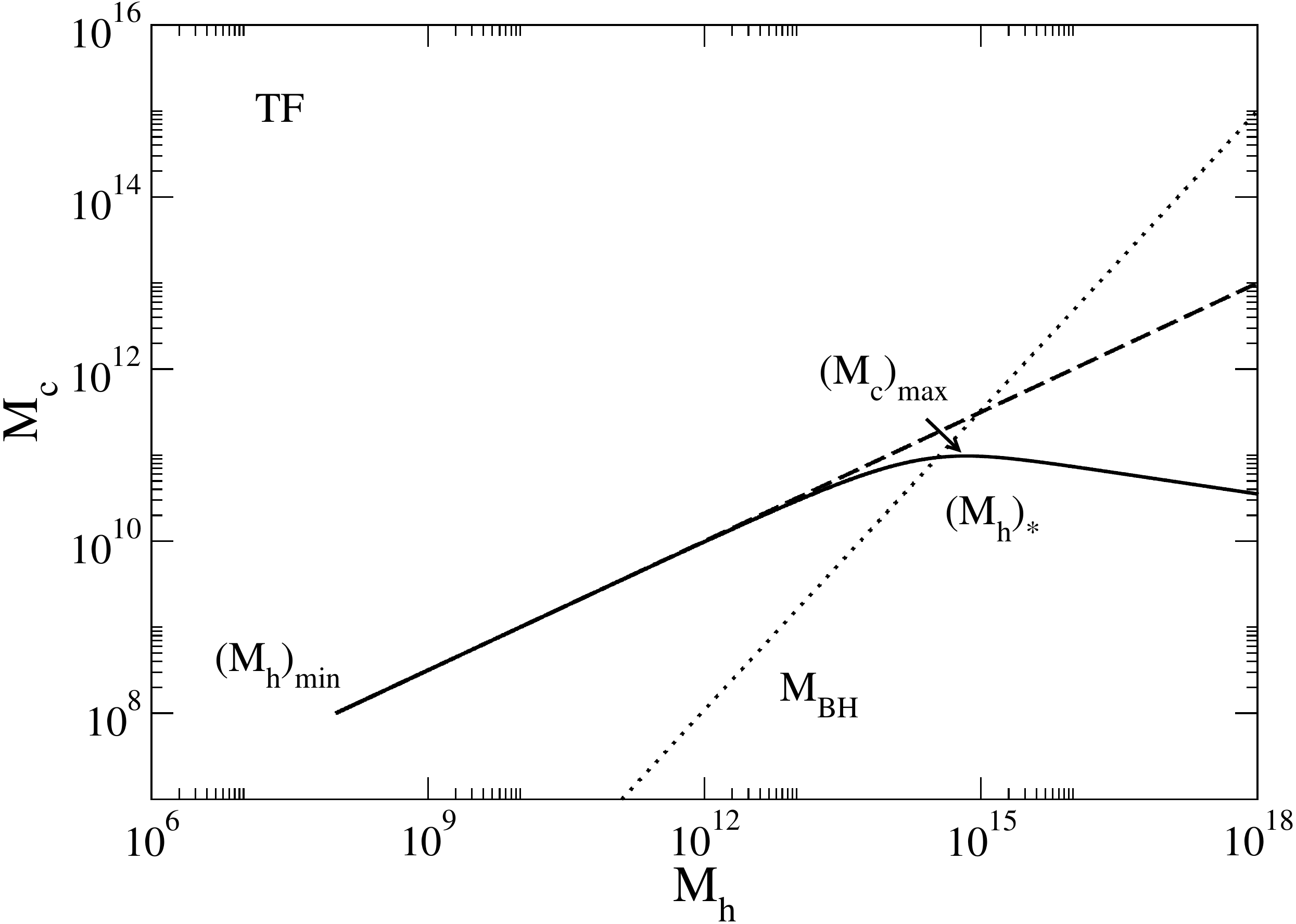}
\caption{Core mass $M_c$ as a function of the halo mass $M_h$ (solid line) for a
repulsive
self-interaction $a_s>0$ in the TF regime. The mass is
normalized by $M_{\odot}$.  The function $M_c(M_h)$ presents a
maximum core mass $(M_c)_{\rm max,TF}=9.77\times 10^{10}\, M_{\odot}$ at
$(M_h)_{\rm *,TF}=6.92\times 10^{14}\, M_{\odot}$. All the configurations are
stable. We have also represented the
relation $M_c(M_h)$ without BH (dashed line) and the
relation $M_{\rm BH}(M_h)$ (dotted line).}
\label{mhmcTF}
\end{center}
\end{figure}

\begin{table*}[t]
\centering
\begin{tabular}{|c|c|c|c|c|c|c|}
\hline
  $\frac{(M_h)_{\rm min}}{(M_h)_{\rm min,0}}$ & $\frac{a_s}{a_*}$ &
$\frac{m}{m_0}$
& $\frac{a_s}{a'_*}$  & $\frac{(M_c)_{\rm max}}{(M_h)_{\rm
min}}$ & $\frac{(M_h)_{*}}{(M_h)_{\rm min}}$\\
\hline
$10^4$ &  $10^4$ & $10^3$ & $10^9$ & $977$ & $6.92\times 10^{6}$
\\
\hline
$1.6$ &  $0.809$ & $1.42$ & $1.46$& $537$ & $4.12\times 10^{6}$
 \\
\hline
$1.1$ &  $0.1465$ & $1.07$ & $0.165$& $167$ & $1.48\times
10^{6}$  \\
\hline
$1.01$ &  $0.015$ & $1.01$ & $0.01515$& $26.8$ & $2.23\times
10^{5}$ \\
\hline
$1$ &  $0$ & $1$ & $0$& $8.92$ & $1.96\times
10^{4}$ \\
\hline
\end{tabular}
\label{table1}
\caption{Values of the DM particle parameters selected in Fig. \ref{mhmc}.
The scales
$m_0$ and $a'_*$ are given by Eqs. (\ref{mh7}) and (\ref{mh8}). For $(M_h)_{\rm
min}=10^8\, M_{\odot}$, we obtain
$m_0=2.25\times 10^{-22}\, {\rm eV}/c^2$ and  $a'_*=4.95\times 10^{-62}\, {\rm
fm}$.}
\end{table*}

We first consider the TF limit  corresponding to $a_s\gg
a_*$. In that case, the core mass -- halo mass
relation (\ref{mcmh6}) reduces to
\begin{eqnarray}
\label{tf1}
&&\frac{M_c}{(M_h)_{\rm min,0}}=-\frac{\lambda}{2\nu}\frac{M_{\rm
BH}}{(M_h)_{\rm
min,0}}\nonumber\\
&+&\sqrt{ \frac{\lambda^2}{4\nu^2}\left (\frac{M_{\rm BH}}{(M_h)_{\rm
min,0}}\right )^2
+ \frac{a_s}{a_*}\frac{M_h}{(M_h)_{\rm min,0}} }.
\end{eqnarray}
Substituting the equivalent
\begin{equation}
\label{tf2}
\frac{(M_{h})_{\rm min}}{(M_{h})_{\rm
min,0}}\sim \frac{a_s}{a_*}
\end{equation}
from Eq. (\ref{mh1}) into Eq. (\ref{tf1}), we obtain
\begin{eqnarray}
\label{tf3}
\frac{M_c}{(M_h)_{\rm min}}&=&-\frac{\lambda}{2\nu}\frac{M_{\rm
BH}}{(M_h)_{\rm
min}}\nonumber\\
&+&\sqrt{ \frac{\lambda^2}{4\nu^2}\left (\frac{M_{\rm BH}}{(M_h)_{\rm
min}}\right )^2
+\frac{M_h}{(M_h)_{\rm min}} },
\end{eqnarray}
where $M_{\rm BH}$ is given as a function of $M_h$  by Eq.
(\ref{bhmhm6}). This relation is
valid for  $a_s\gg
a'_*$. It is plotted in
Fig. \ref{mhmcTF}. There is a maximum core mass
\begin{eqnarray}
\label{tf4}
(M_c)_{\rm max,TF}=9.77\times 10^{10}\, M_{\odot},
\end{eqnarray}
corresponding to a halo mass
\begin{eqnarray}
\label{tf5}
(M_h)_{\rm *,TF}=6.92\times 10^{14}\, M_{\odot}
\end{eqnarray}
and a BH mass
\begin{eqnarray}
\label{tf6}
(M_{\rm BH})_{\rm *,TF}=2.16\times 10^{11}\, M_{\odot}.
\end{eqnarray}
The effect of the BH becomes
important when $M_{\rm BH}/(M_h)_{\rm min}\sim ({M_h}/(M_h)_{\rm min})^{1/2}$,
i.e. $M_{\rm BH}\sim M_c$, corresponding  to $(M_h)_{\rm *,TF}\sim 10^{15}\,
M_{\odot}$.\footnote{Since the value of $(M_h)_{\rm *,TF}$ is larger than
the size of the biggest DM halos in the Universe, we conclude that the effect of
the BH is always negligible (or marginal for the biggest halos) when we are in
the TF limit.} When $M_h\ll (M_h)_{\rm *,TF}$ the
effect of the BH is
negligible we recover the scaling 
\begin{eqnarray}
\label{tf7}
\frac{M_c}{(M_h)_{\rm min}}=\left
(\frac{M_h}{(M_h)_{\rm min}}\right
)^{1/2}
\end{eqnarray}
obtained in \cite{mcmh}. When $M_h\gg (M_h)_{\rm *,TF}$, the BH dominates and
we 
get the
scaling
\begin{eqnarray}
\label{tf8}
\frac{M_c}{(M_h)_{\rm min}}&\sim& \frac{\nu}{\lambda}\frac{M_h}{(M_h)_{\rm
min}}\frac{(M_h)_{\rm min}}{M_{\rm
BH}}\nonumber\\
&\sim& \frac{\nu}{\lambda A}\left
(\frac{(M_h)_{\rm min}}{M_h}\right )^{a-1}.
\end{eqnarray}
This relation can be directly obtained from Eqs. (\ref{mcmh2})
and (\ref{mrr7q}). It exhibits a critical index $a_{\rm TF}=1$. When $a>a_{\rm
TF}$ the
core mass
decreases with $M_h$ and when  $a<a_{\rm TF}$ the core mass
increases with $M_h$. For the measured value $a=1.16$ we are in the first case.
But since $a=1.16$ is close to the critical value $a_{\rm TF}=1$, the decrease
of the core mass $M_c$ with $M_h$ is
weak, as can be seen in Fig. \ref{mhmcTF}. For a DM halo of
mass
$M_h=10^{12}\,
M_{\odot}$ similar to the one that surrounds our Galaxy, we obtain a core
mass $M_c=10^{10}\,
M_{\odot}$, the same value as the one found in \cite{mcmh} in the absence of a
central BH (we have taken $(M_h)_{\rm min}=10^{8}\,
M_{\odot}$).

\begin{figure}[!h]
\begin{center}
\includegraphics[clip,scale=0.3]{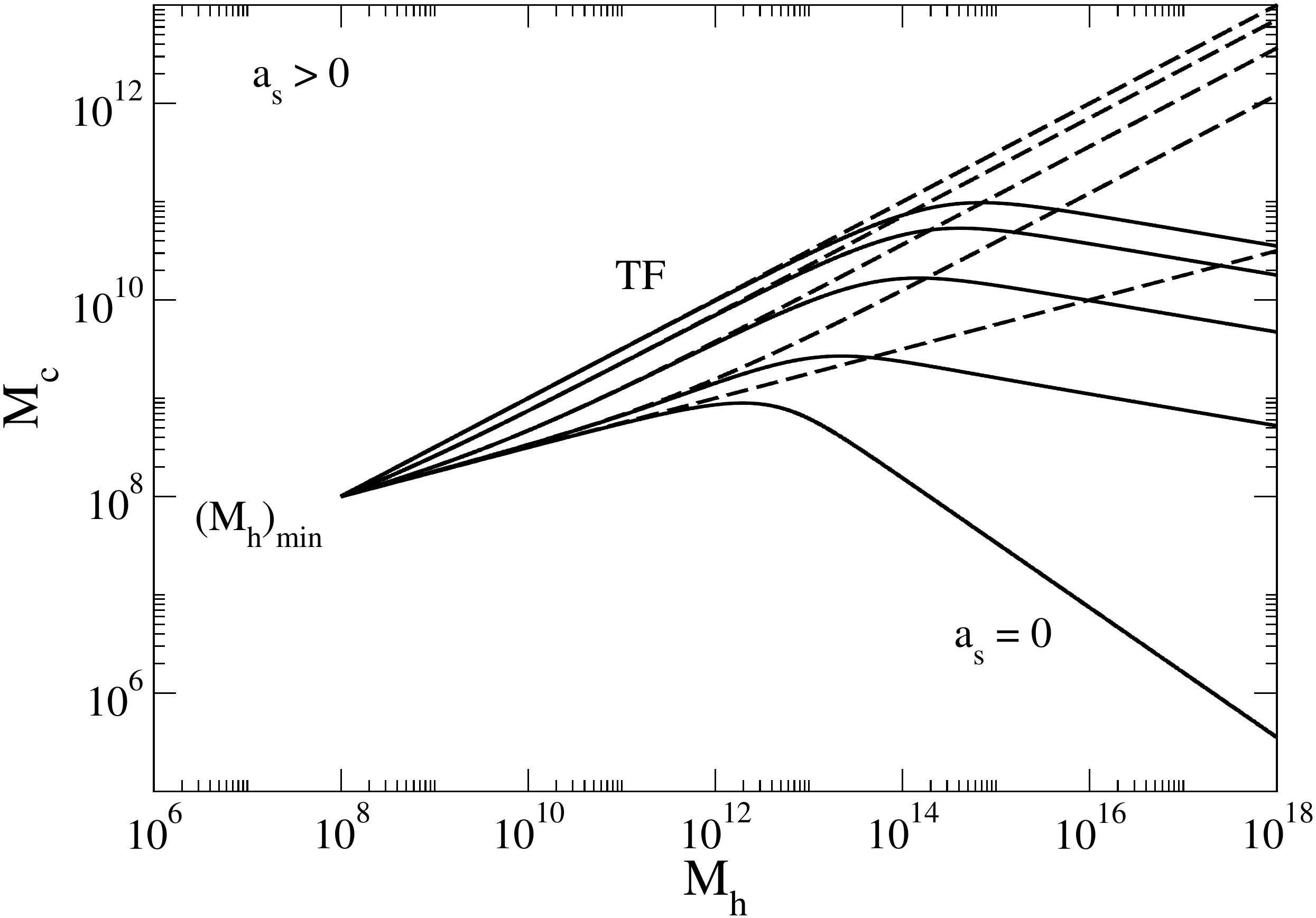}
\caption{Core mass $M_c$ as a function of the halo mass $M_h$ for different
values of the scattering length $a_s\ge 0$ (see Table I). The
mass is
normalized by $M_{\odot}$. We have represented
the position of the minimum halo mass $(M_{h})_{\rm min}=10^8\,
M_{\odot}$ (common origin), the curve corresponding to noninteracting bosons
$a_s=0$ [see Eq. (\ref{ni1})], and the curve corresponding to the
TF limit
$a_s/a'_*\gg 1$ [see Eq. (\ref{tf3})]. The core mass reaches a 
maximum value $(M_c)_{\rm max}$ at  $(M_{h})_{*}$. The dashed line represents
the core mass -- halo mass relation $M_c(M_h)$ in the absence of a central BH
\cite{mcmh}. In that case, the core mass increases monotonically with the halo
mass. }
\label{mhmc}
\end{center}
\end{figure}

We now consider the general case of a repulsive self-interaction ($a_s>0$). The
core mass -- halo mass relation $M_c(M_h)$ is represented in Fig. \ref{mhmc} for
different values of $a_s$ following the procedure explained in Sec.
\ref{sec_proc}. These curves are bounded by the
lower curve from Eq. (\ref{ni1}) corresponding to the noninteracting limit
($a_s\ll a'_*$) and by the upper curve from Eq. (\ref{tf3}) corresponding to the
TF limit ($a_s\gg a'_*$). The general behavior of the function $M_c(M_h)$ is
always the same. Starting from the minimum halo mass $(M_h)_{\rm min}$ where
$M_c=M_h$, the core mass $M_c$ increases with $M_h$, reaches a maximum
$(M_c)_{\rm max}$ at $(M_h)_{*}$, and finally decreases with $M_h$. All the
halos are stable. We see that the maximum mass $(M_c)_{\rm max}$ increases with
$a_s$ going from $(M_c)_{\rm max,0}=8.92\times 10^8\,
M_{\odot}$ for $a_s=0$ to $(M_c)_{\rm max,TF}=9.77\times 10^{10}\, M_{\odot}$ 
for $a_s/a'_*\gg 1$. The corresponding halo mass $(M_h)_{*}$ increases with
$a_s$ going from $(M_h)_{*,0}=1.96\times 10^{12}\,
M_{\odot}$ to $(M_h)_{\rm *,TF}=6.92\times 10^{14}\, M_{\odot}$. The effect
of the BH is negligible for small halo masses $M_h\ll (M_h)_{*}$. It becomes
important when $M_h\sim (M_h)_{*}$.

\subsection{Bosons with an attractive self-interaction}
\label{sec_att}

In this section, we consider the case of bosons with an attractive
self-interaction ($a_s<0$).

According to Eq. (\ref{mcmh6}) the core mass  vanishes at
\begin{equation}
\label{att1}
\frac{(M_h)_{\rm Max}}{(M_h)_{\rm min,0}}=\left
(\frac{a_*}{a_s}\right )^2,
\end{equation}
like in the absence of a central BH \cite{mcmh}.
However, this situation is purely academic
because, as we shall see, the core becomes unstable long before disappearing.

\begin{figure}[!h]
\begin{center}
\includegraphics[clip,scale=0.3]{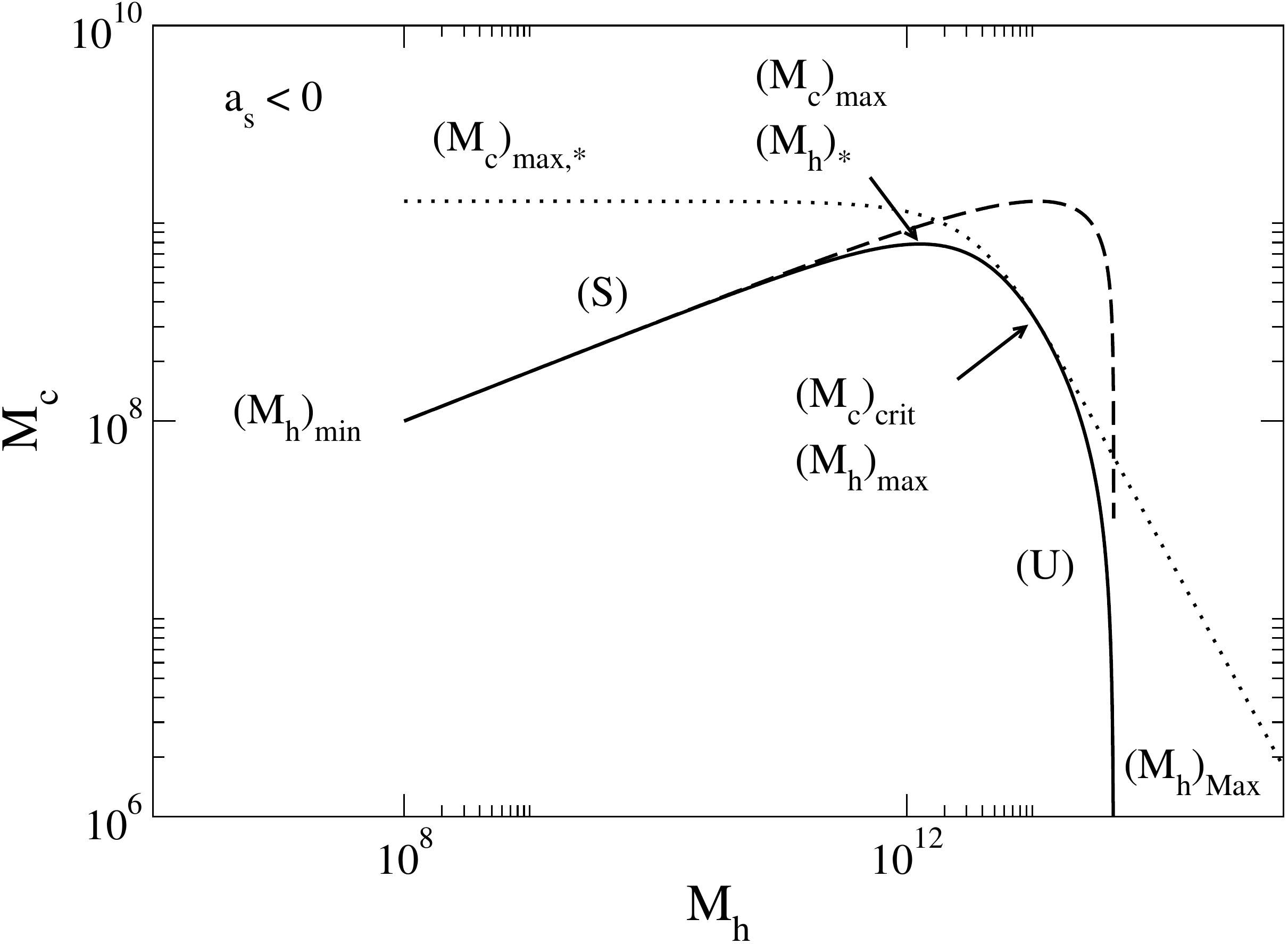}
\caption{Core mass $M_c$ as a function of the halo mass $M_h$ for a
specific value of $a_s<0$ (we have taken $a_s/a'_*=-0.00150$). 
The core mass reaches a  maximum value $(M_c)_{\rm max}$ at  $(M_{h})_{*}$.
However, the DM halo is stable until the maximum halo mass
$(M_{h})_{\rm max}$ corresponding to a critical core mass $(M_c)_{\rm crit}$
[see Eq. (\ref{mrr8})].
The dashed
line represents the core mass -- halo mass relation $M_c(M_h)$ without a central
BH. The dotted line represents the maximum mass of the core in
the presence of
a BH of mass $M_{\rm BH}(M_h)$.}
\label{mhmcATTexemple}
\end{center}
\end{figure}

The core mass -- halo mass
relation $M_c(M_h)$ is
plotted in Fig. \ref{mhmcATTexemple} for a given value of $a_s<0$.  The function
$M_c(M_h)$ presents a maximum $(M_c)_{\rm max}$ at $(M_h)_{*}$.
It is important to note that
the maximum of the function $M_c(M_h)$ does {\it not} generally coincide with
the maximum mass
$(M_c)_{\rm max,*}$ of the quantum core in the presence of a BH of mass $M_{\rm
BH}$ [see Eq. (\ref{mrr8})]. The evolution of the critical mass $(M_c)_{\rm
max,*}$ with $M_h$ is
determined by the equation\footnote{This equation is obtained by
combining Eq. (\ref{mrr8}) with the relation
\begin{equation}
\label{att3b}
\frac{(M_c)_{\rm max}}{(M_h)_{\rm min,0}}=\frac{1}{2}\left
(\frac{a_*}{|a_s|}\right )^{1/2}
\end{equation}
given in \cite{mcmh} in the absence of a central BH.}
\begin{eqnarray}
\label{att3}
\frac{(M_c)_{\rm max,*}}{(M_h)_{\rm min,0}}&=&-\frac{\lambda}{2\nu}\frac{M_{\rm
BH}}{(M_h)_{\rm
min,0}}\nonumber\\
&+&\sqrt{ \frac{\lambda^2}{4\nu^2}\left (\frac{M_{\rm BH}}{(M_h)_{\rm
min,0}}\right )^2
+\frac{1}{4}\frac{a_*}{|a_s|} },
\end{eqnarray}
together with Eq. (\ref{bhmhm6}). The
intersection between the curves $(M_c)(M_h)$ and 
$(M_c)_{\rm max,*}(M_h)$ determines the point $((M_h)_{\rm
max},(M_c)_{\rm crit})$ at which the core of the halo becomes
unstable (it is marginally stable at that point). The branch
$(M_h)_{\rm max}\le M_h\le (M_h)_{\rm
Max}$ corresponds to unstable states. Only the branch
$(M_h)_{\rm
min}\le M_h\le (M_h)_{\rm
max}$, corresponding to stable states, is physical. Therefore,
$(M_h)_{\rm
max}$ is the maximum mass of a DM halo with a stable quantum
core.

\begin{figure}[!h]
\begin{center}
\includegraphics[clip,scale=0.3]{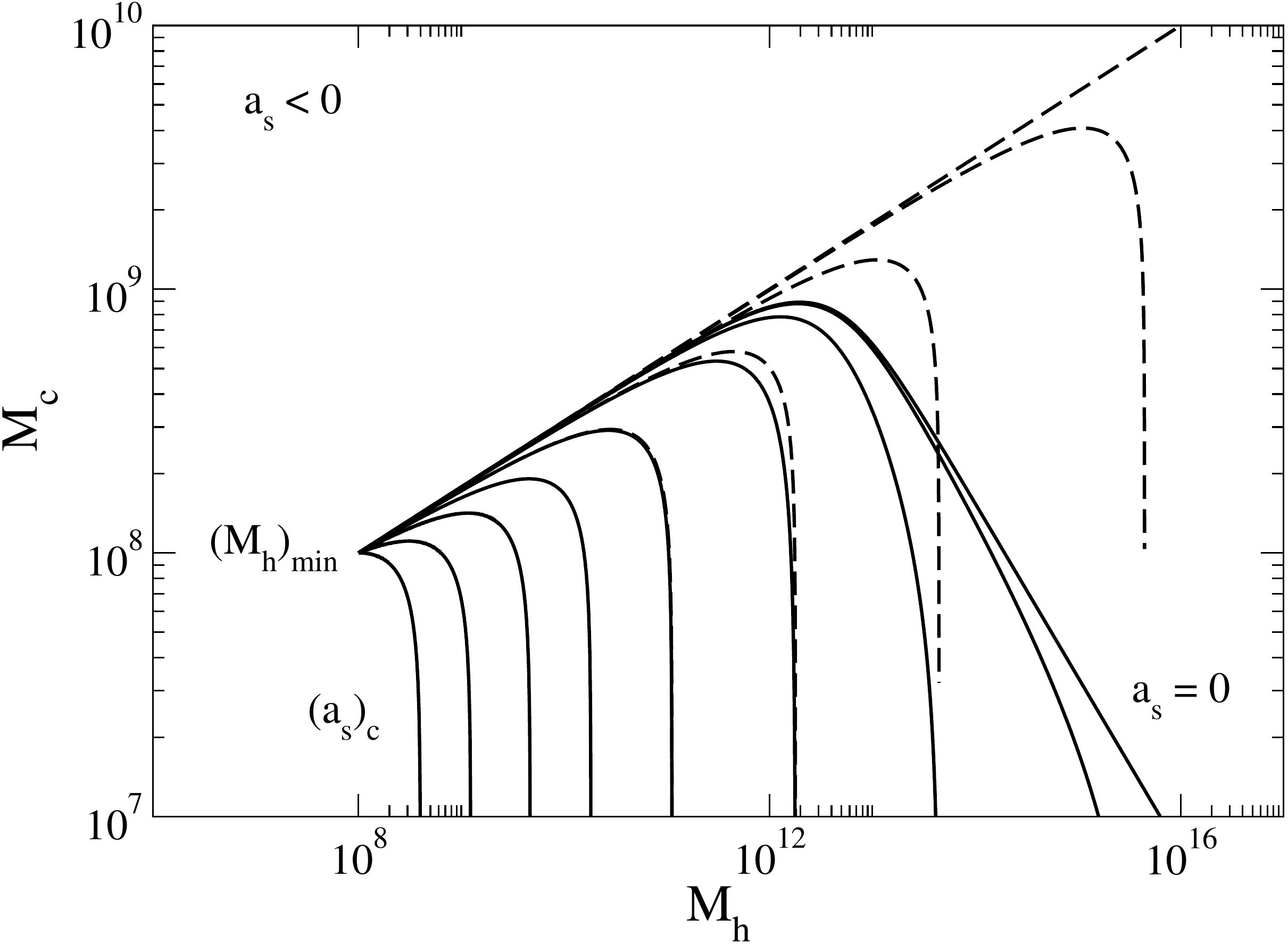}
\caption{Core mass $M_c$ as a function of the halo mass $M_h$ for different
values of the scattering length $(a_s)_c\le a_s\le 0$ (see Table II). The
mass is normalized by $M_{\odot}$. We have represented
the position of the minimum halo mass $(M_{h})_{\rm min}=10^8\,
M_{\odot}$ (common origin), the curve corresponding the minimum scattering
length $(a_s)_c$ for which the minimum halo is critical, and the
curve corresponding to noninteracting bosons $a_s=0$ [see Eq. (\ref{ni1})]. The
core mass reaches a 
maximum value $(M_c)_{\rm max}$ at  $(M_{h})_{*}$. The dashed lines represent
the core mass -- halo mass relation $M_c(M_h)$ in the absence of a central BH
\cite{mcmh}.}
\label{mhmcATT}
\end{center}
\end{figure}

\begin{figure}[!h]
\begin{center}
\includegraphics[clip,scale=0.3]{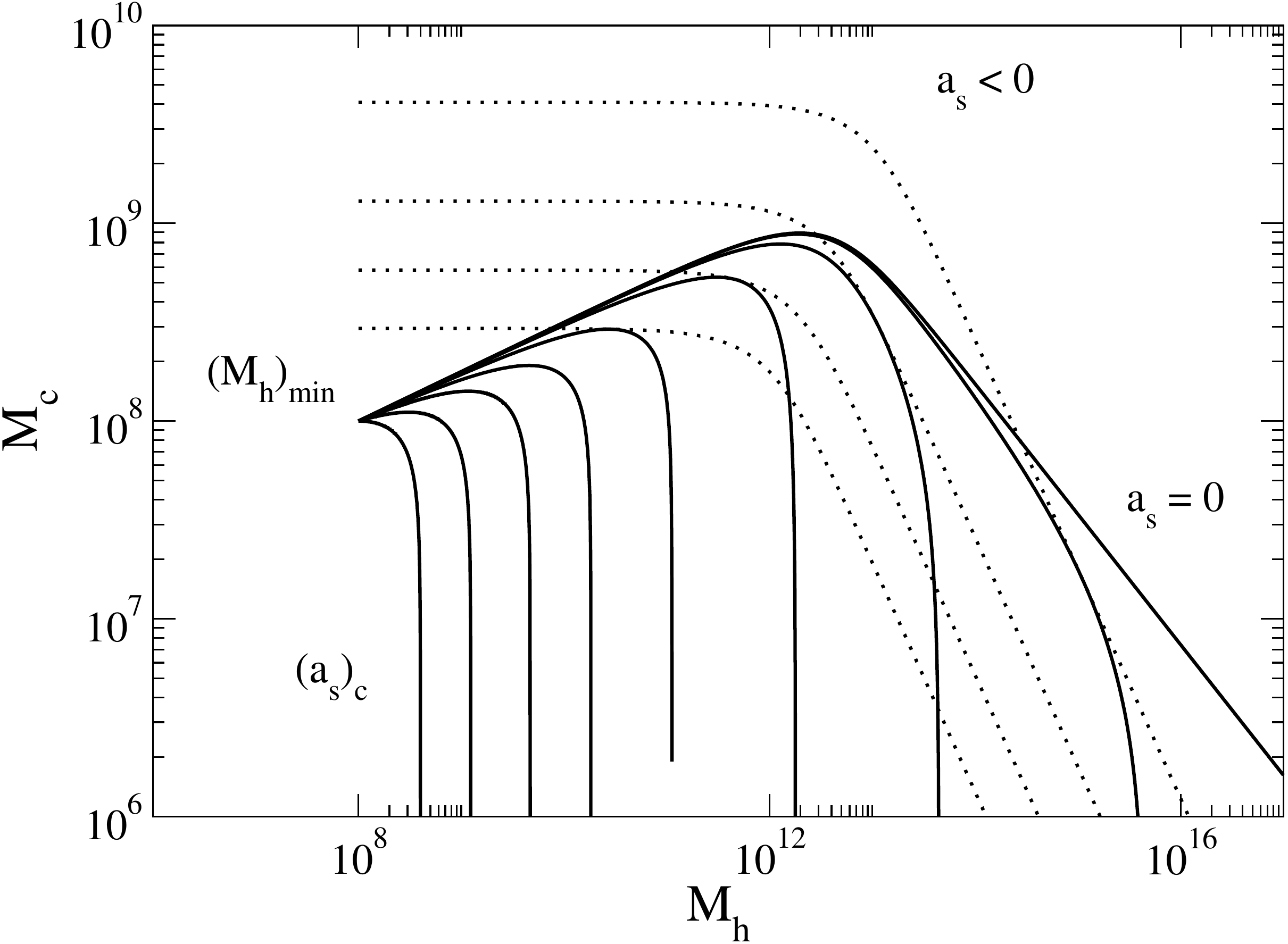}
\caption{Same as Fig. \ref{mhmcATT} but now the dotted lines represent the
maximum mass of the quantum core  $(M_c)_{\rm max,*}(M_h)$ in the presence of a
BH of
mass $M_{\rm BH}(M_h)$. Their intersections with the core mass -- halo mass
relations $M_c(M_h)$ (solid lines) determine the points $((M_h)_{\rm
max},(M_c)_{\rm crit})$ at which the core of the halo becomes
unstable. For values of $a_s$ close to  $(a_s)_c$  we have not
plotted
the dotted lines because $(M_c)_{\rm max}$ and $(M_c)_{\rm crit}$ coincide. When
$a_s<(a_s)_c$ (i.e. $f<f_c$) there is no DM halo with a stable quantum core.
When
$(a_s)_c<a_s<0$ a stable quantum core exists
only in the range $(M_h)_{\rm min}\le M_h\le (M_h)_{\rm max}$.}
\label{mhmcATTstab}
\end{center}
\end{figure}

We now consider the general case of an attractive self-interaction ($a_s<0$).
The core
mass -- halo mass relation $M_c(M_h)$ is represented in Figs. \ref{mhmcATT} and
\ref{mhmcATTstab} for different values
of $a_s$ in the range $[(a_s)_c,0]$ following the procedure explained in Sec.
\ref{sec_proc}. These curves are bounded by  the
lower
curve $(M_h)_{\rm min}=(M_h)_{\rm max}$ (critical minimum halo)
corresponding to the minimum scattering length $(a_s)_c$ and
by the upper curve
from Eq.  (\ref{ni1})
corresponding to the noninteracting limit  ($|a_s|\ll a'_*$). The general
behavior of the
$M_c(M_h)$ relation is always the same. Starting from the minimum halo mass
$(M_h)_{\rm min}$ where $M_c=M_h$, the core mass $M_c$ increases with $M_h$,
reaches a
maximum at $(M_c)_{\rm max}$ at $(M_h)_{*}$ and finally decreases with $M_h$.
The halos
are stable until $(M_h)_{\rm max}$ corresponding to $M_c=(M_c)_{\rm crit}$. For
$M_h>(M_h)_{\rm max}$ they are unstable. For  $M_h=(M_h)_{\rm
Max}$ the core mass vanishes ($M_c=0$).  We see that
the maximum mass $(M_c)_{\rm max}$ increases with $a_s$, going from $(M_h)_{\rm
min}=10^8\, M_{\odot}$ for
$a_s=(a_s)_c$ to $(M_c)_{\rm max,0}=8.92\times 10^8\,
M_{\odot}$  for $a_s=0$. The corresponding halo
mass $(M_h)_{*}$ increases with $a_s$  going
from $(M_h)_{\rm
min}=10^8\, M_{\odot}$ to $(M_h)_{*,0}=1.96\times 10^{12}\,
M_{\odot}$.\footnote{This is very different from the case without a central BH
where the maximum core mass and the  corresponding halo
mass increase to infinity when $a_s$ goes to $0^-$ \cite{mcmh}.} On the other
hand, the critical mass $(M_c)_{\rm crit}$
decreases with $a_s$ going from $(M_h)_{\rm
min}=10^8\, M_{\odot}$ for
$a_s=(a_s)_c$ to $0$ for $a_s=0$ while the maximum halo mass $(M_h)_{\rm
max}$ increases with
$a_s$ going from $(M_h)_{\rm
min}=10^8\, M_{\odot}$ to $+\infty$. 
The effect of the BH is negligible for $a_s$ close to
$(a_s)_c$. In that case $(M_c)_{\rm crit}\simeq (M_c)_{\rm max}$ and
$(M_h)_{\rm max}\simeq (M_h)_{*}$. It is also negligible for any $a_s$ when
$M_h\ll (M_h)_*$. It becomes
important when  $a_s$ is close
to $0$ and $M_h\sim (M_h)_*$.

\begin{table*}[t]
\centering
\begin{tabular}{|c|c|c|c|c|c|c|c|c|}
\hline
  $\frac{(M_h)_{\rm min}}{(M_h)_{\rm min,0}}$ & $\frac{a_s}{a_*}$ &
$\frac{m}{m_0}$
& $\frac{a_s}{a'_*}$ & $\frac{f}{f'_*}$ & $\frac{(M_h)_{*}}{(M_h)_{\rm
min}}$ & $\frac{(M_c)_{\rm max}}{(M_h)_{\rm min}}$ &
$\frac{(M_h)_{\rm max}}{(M_h)_{\rm min}}$ &  $\frac{(M_c)_{\rm crit}}{(M_h)_{\rm
min}}$\\
\hline
$1$ &  $0$ & $1$ & $0$& $\infty$ & $1.96\times 10^4$ &  $8.92$ & $\infty$ &
$0$\\
\hline
$0.9999$ &  $-1.50\times 10^{-4}$ & $1.00$ & $-1.50\times 10^{-4}$ & $81.6$ &
$1.88\times 10^4$ & $8.81$ &
$1.11\times 10^7$ & $0.158$ \\
\hline
$0.999$ &  $-1.50\times 10^{-3}$ & $0.999$ & $-0.00150$ & $25.8$ & $1.27\times
10^4$ & $7.85$ & $1.11\times 10^5$ & $3.11$ \\
\hline
$0.995$ &  $-0.00751$ & $0.996$ & $-0.00746$ & $11.55$ & $3040$ & $5.33$ &
$4460$ & $5.22$ \\
\hline
$0.98$ &  $-0.03015$ & $0.985$ & $-0.0294$ & $5.79$ & $270$  & $2.91$ & $281$ &
$2.91$ \\
\hline
$0.95$ &  $-0.0760$ & $0.962$ & $-0.0713$ & $3.67$ & $45.6$ & $1.91$ & $45.6$
& $1.91$ \\
\hline
$0.9$ &  $-0.154$ & $0.924$ & $-0.135$ & $2.615$ & $11.7$ & $1.41$ & $11.7$ &
$1.41$\\
\hline
$0.8$ &  $-0.318$ & $0.846$ & $-0.241$ & $1.87$ & $3.09$ & $1.11$ & $3.09$ &
$1.11$\\
\hline
$0.630$ &  $-0.630$ & $0.707$ & $-0.354$ & $1.41$ & $1$ & $1$ & $1$ & $1$ \\
\hline
\end{tabular}
\label{table2}
\caption{Values of the DM particle parameters selected in Figs.
\ref{mhmcATT} and \ref{mhmcATTstab}. The scales
$m_0$, $a'_*$ and $f'_*$ are given by Eqs. (\ref{mh7}), (\ref{mh8}) 
and (\ref{mh14b}). For $(M_h)_{\rm min}=10^8\, M_{\odot}$, we obtain
$m_0=2.25\times 10^{-22}\, {\rm eV}/c^2$, $a'_*=4.95\times 10^{-62}\, {\rm fm}$
and $f'_*=9.45\times 10^{13}\, {\rm GeV}$.}
\end{table*}

\subsection{Summary}
\label{sec_summary}

In the noninteracting case ($a_s=0$; $m=m_0$ given by Eq. (\ref{mh9})) the
DM halos
with a mass $M_h\ge (M_h)_{\rm
min}$ [see Eq. (\ref{mh5})] contain a quantum core of mass $M_c$ given
by Eq.
(\ref{ni1}). The core mass achieves a maximum value $(M_c)_{\rm max,0}$ at
$(M_h)_{*,0}$
[see Eqs. (\ref{ni2}) and (\ref{ni3})]. All the configurations are stable.

In the case of a repulsive self-interaction ($a_s>0$; $m>m_0$ given by Eq.
(\ref{mh6}))
the halos with a mass
$M_h>(M_h)_{\rm min}$ [see Eq. (\ref{mh5})] contain a quantum core of mass
$M_c$ given by Eq. (\ref{mcmh6}).  The core mass achieves a maximum value
$(M_c)_{\rm
max}$ at $(M_h)_{*}$. For $a_s\ll a_*$,
we are in the noninteracting limit discussed previously. For $a_s\gg a_*$  we
are in the
TF limit. In that case,  the  mass $M_c$ of the quantum core is given by Eq.
(\ref{tf3}).  The core mass achieves a maximum value $(M_c)_{\rm max,TF}$ at
$(M_h)_{\rm *,TF}$
[see Eqs. (\ref{tf4}) and (\ref{tf5})].  All the configurations are stable.

In the case of an attractive self-interaction ($a_s<0$) the
halos can be stable only if $a_s>(a_s)_c$ (or equivalently $f>f_c$). When
$(a_s)_c\le a_s<0$ (correspondingly $m_c\le
m<m_0$ given by Eq. (\ref{mh6})) the halos with a mass
$M_h>(M_h)_{\rm min}$ [see Eq. (\ref{mh5})] contain a quantum core of mass
$M_c$ given by Eq. (\ref{mcmh6}). The core mass achieves a maximum value
$(M_c)_{\rm
max}$ at $(M_h)_{*}$. The quantum core is stable for $(M_h)_{\rm
min}<M_h<(M_h)_{\rm
max}$ and unstable for $(M_h)_{\rm
max}<M_h<(M_h)_{\rm
Max}$. When we reach the
maximum halo mass $(M_h)_{\rm max}$, the core
reaches
its maximum limit $(M_c)_{\rm crit}$ [see Eq.
(\ref{mrr8})] and collapses. The
result of the
collapse (dense
axion star, BH, bosenova...) is discussed
in \cite{braaten,cotner,bectcoll,ebycollapse,tkachevprl,helfer,phi6,visinelli,
moss}.

\begin{figure}[!h]
\begin{center}
\includegraphics[clip,scale=0.3]{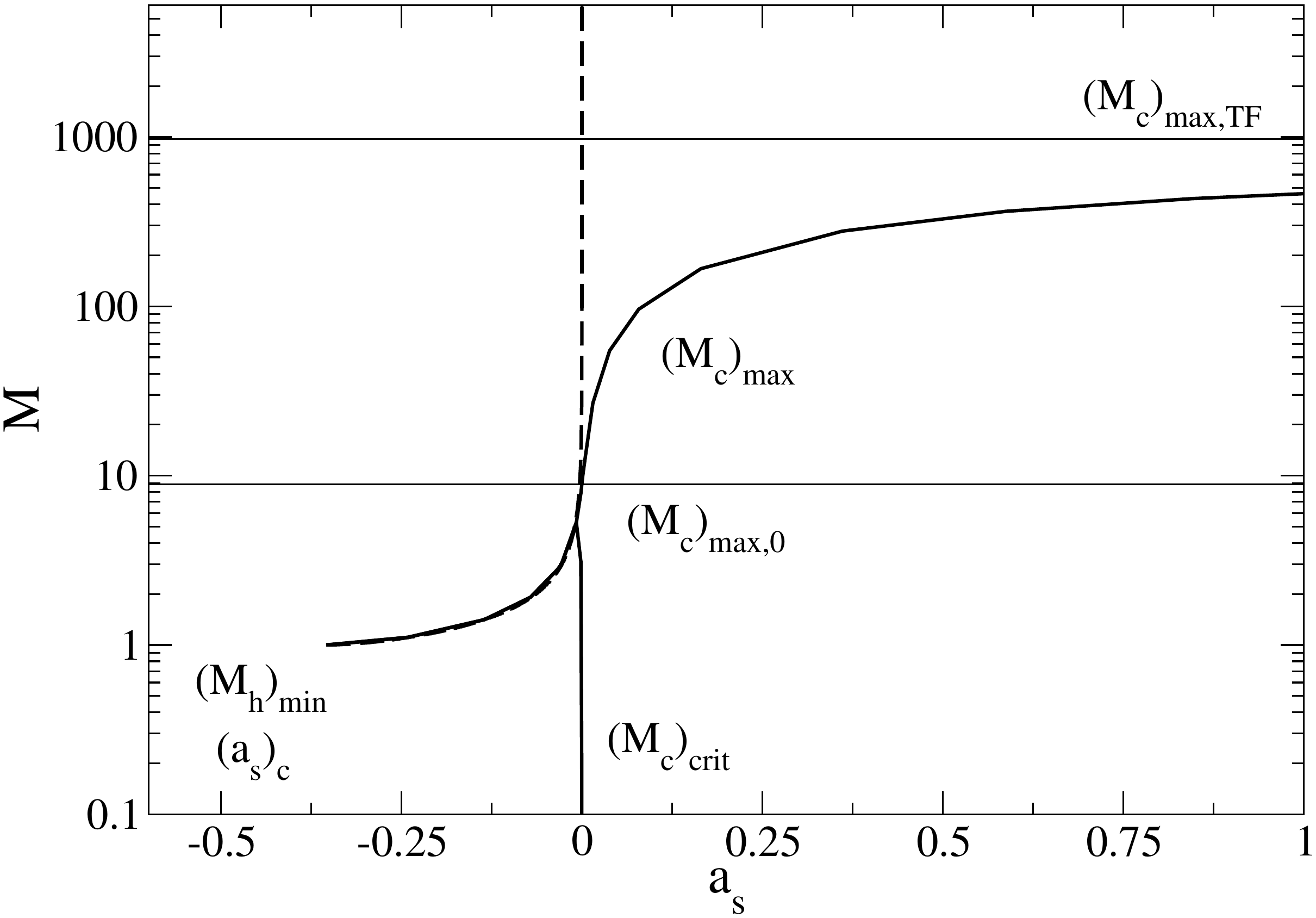}
\caption{Maximum core mass $(M_c)_{\rm max}/(M_h)_{\rm min}$ and critical core
mass $(M_c)_{\rm crit}/(M_h)_{\rm min}$  as a function of the scattering length
$a_s/a'_*$. For $a_s\ge 0$, the critical core mass  $(M_c)_{\rm
crit}=0$ meaning that all the configurations are stable. The dashed line
represents the maximum core mass $(M_c)_{\rm max}/(M_h)_{\rm min}$ in the
absence of a central BH [see Eq. (239) of \cite{mcmh}] which tends to $+\infty$
as
$a_s\rightarrow 0^-$. }
\label{asMc}
\end{center}
\end{figure}

\begin{figure}[!h]
\begin{center}
\includegraphics[clip,scale=0.3]{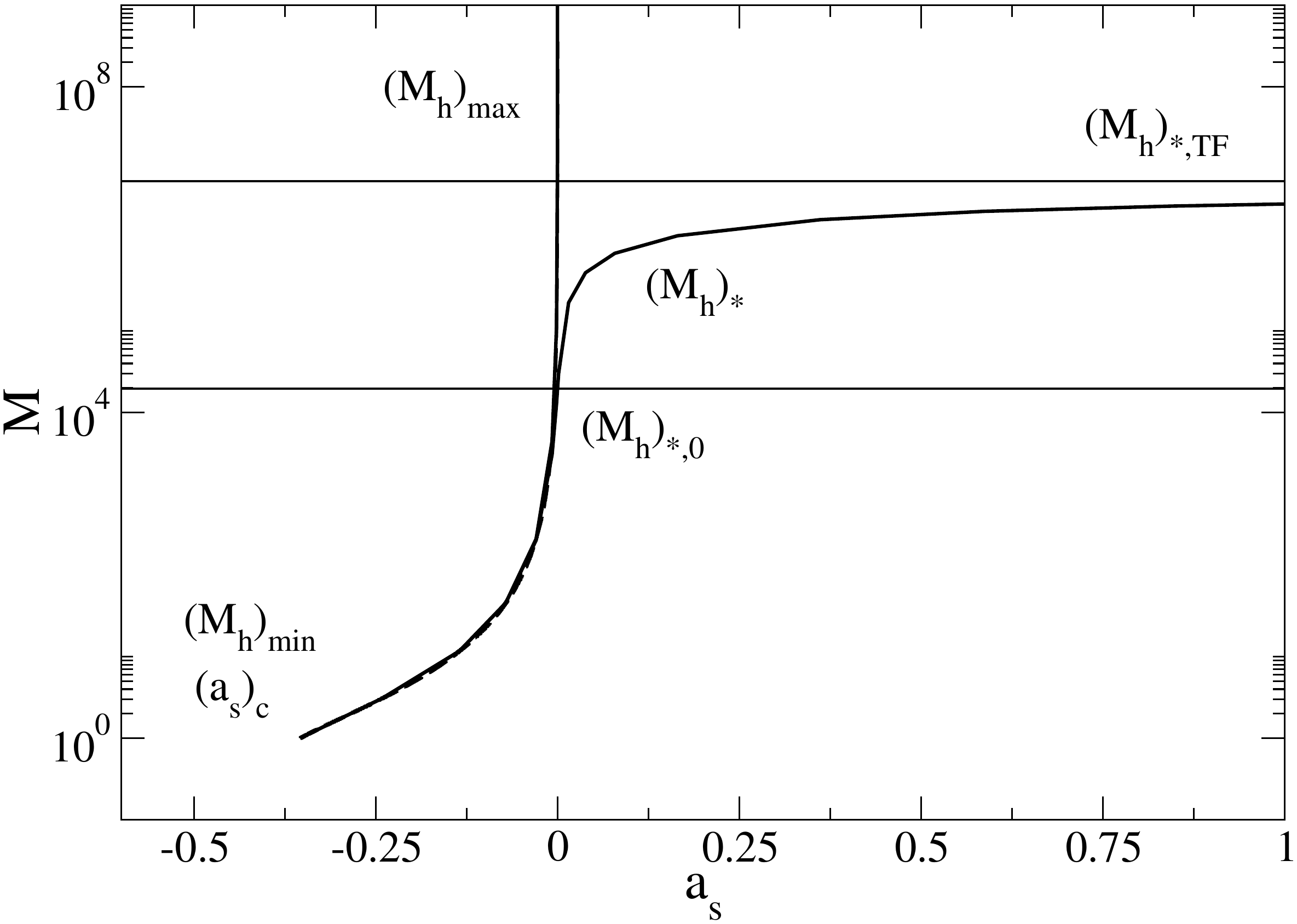}
\caption{Halo mass $(M_h)_{*}/(M_h)_{\rm min}$ corresponding to the maximum
core mass and maximum halo mass  $(M_h)_{\rm max}/(M_h)_{\rm min}$ above which
the quantum core is unstable as a function of the scattering length
$a_s/a'_*$. For $a_s\ge 0$, the maximum halo mass  $(M_h)_{\rm max}=+\infty$
meaning that all the configurations are stable. The dashed line
represents the maximum halo mass $(M_h)_{\rm max}/(M_h)_{\rm min}$ in
the absence of a central BH [see Eq. (238) of \cite{mcmh}].}
\label{asMh}
\end{center}
\end{figure}

The maximum core mass $(M_c)_{\rm max}$ and the critical core
mass $(M_c)_{\rm crit}$  are plotted as a
function of $a_s$ in Fig. \ref{asMc}. The corresponding halo masses
$(M_h)_{*}$ and  $(M_h)_{\rm max}$ are plotted as a
function of $a_s$ in Fig. \ref{asMh}.

In the previous sections, we have expressed the core mass -- halo mass relation
in terms of $M_h$. This relation  can be easily expressed in terms of $M_v$
by using Eq. (\ref{bhmhm2b}).

\section{Fermionic DM halos}
\label{sec_fdm}

In this section, we obtain the core mass -- halo mass relation of fermionic DM
halos in the presence of a central BH by combining the velocity
dispersion tracing relation (\ref{mcmh2}) with the core mass-radius
relation of a self-gravitating gas of fermions at $T=0$.

\subsection{Core mass-radius relation}
\label{sec_fmrr}

Using a Gaussian ansatz, we found in \cite{epjpbh} that the approximate
mass-radius relation of a self-gravitating gas of fermions at $T=0$ (ground
state) with a
central BH is given
by
\begin{equation}
\label{fmrr1}
R_c=\frac{3\zeta}{20}\left (\frac{3}{\pi}\right
)^{2/3}\frac{h^2}{Gm^{8/3}}\frac{M_c^{2/3}}{\nu M_c+\lambda M_{\rm BH}}
\end{equation}
with the coefficients $\zeta=1/(2\pi)^{3/2}$,
$\nu=1/\sqrt{2\pi}$ and $\lambda=2/\sqrt{\pi}$. These
results apply to the  ``minimum halo''  which has no isothermal
atmosphere (ground
state) and to the
quantum core of larger DM halos which have an isothermal atmosphere. Starting
from $R_c=0$ when $M_c=0$, the radius $R_c$ increases as the mass $M_c$
increases,
reaches a maximum
\begin{eqnarray}
\label{fmrr3}
(R_{\rm c})_{\rm max}(M_{\rm BH})=\frac{\zeta}{20\lambda^{1/3}}\left
(\frac{6}{\pi\nu}\right )^{2/3} \frac{h^2}{Gm^{8/3}M_{\rm BH}^{1/3}}
\end{eqnarray}
at
\begin{eqnarray}
\label{fmrr4}
M_*(M_{\rm BH})=\frac{2\lambda}{\nu}M_{\rm BH}
\end{eqnarray}
and decreases to zero  when  $M\rightarrow +\infty$ (see Fig. 18 of
\cite{epjpbh}). All these configurations are stable.
For a given value of the core mass $M_c$, the core radius decreases as
the BH mass increases going from 
\begin{equation}
\label{fmrr2}
R_c=\frac{3\zeta}{20\nu}\left (\frac{3}{\pi}\right
)^{2/3}\frac{h^2}{Gm^{8/3}M_c^{1/3}}
\end{equation}
when $M_{\rm BH}=0$ to  
\begin{equation}
\label{fmrr1q}
R_c\sim \frac{3\zeta}{20\lambda}\left (\frac{3}{\pi}\right
)^{2/3}\frac{h^2 M_c^{2/3}}{Gm^{8/3}M_{\rm
BH}}\rightarrow 0
\end{equation}
when the BH dominates ($M_{\rm BH}\rightarrow +\infty$). Although the central
BH
enhances the
gravitational attraction and reduces the radius of the BEC, it
does not destabilize the system.

\subsection{The minimum halo mass without central BH}
\label{sec_fmhm}

We first determine the minimum halo mass $(M_h)_{\rm min}$ in the absence of a
central BH. As explained
previously, the minimum halo corresponds to the
ground state ($T=0$) of the self-gravitating Fermi gas. In our
approximate approach
we
write the surface density as
\begin{equation}
\label{fmrr5}
\Sigma_0=\alpha\frac{M_c}{R_c^2},
\end{equation}
where $\alpha$ is a constant of order unity (in the numerical applications we
take $\alpha=1/1.76$ for the reason explained in footnote 17 of \cite{mcmh}).
Eliminating
$R_c$ between
Eqs. (\ref{fmrr2}) and (\ref{fmrr5}), and treating $\Sigma_0$ as a universal
constant,
we get the minimum halo mass $(M_{h})_{\rm min}$ as a function of $m$
(we recall that $M_h=M_c$ for the ground state since there is no isothermal
atmosphere by definition). We find 
\begin{equation}
\label{fmrr6}
(M_h)_{\rm min}=\frac{(2\pi)^{12/5}}{\alpha^{3/5}\nu^{6/5}}\left
(\frac{3\zeta}{20}\right
)^{6/5}\left (\frac{3}{\pi}\right )^{4/5}\left
(\frac{\hbar^{12}\Sigma_0^3}{G^6m^{16}}\right )^{1/5}.
\end{equation}
The prefactor is $1.26$. This result
can be compared with the exact value from Eq.
(14) of \cite{mcmh}. Measuring the DM particle  mass in units of
$100 \, {\rm eV/c^2}$, we get $(M_h)_{\rm min}=1.40\times 10^8 m^{-16/5}\,
M_{\odot}$.
Inversely, assuming that the mass $(M_h)_{\rm
min}$ of the minimum halo is known, we obtain the fermion mass
\begin{equation}
\label{fmrr7}
m=\frac{(2\pi)^{3/4}}{\alpha^{3/16}\nu^{3/8}}\left (\frac{3\zeta}{20}\right
)^{3/8}\left
(\frac{3}{\pi}\right
)^{1/4}\frac{\hbar^{3/4}\Sigma_0^{3/16}}{G^{3/8}(M_h)_{\rm
min}^{5/16}}.
\end{equation}
The prefactor is $1.07$. If we take
$(M_h)_{\rm min}=10^8\, M_{\odot}$ we obtain $m=111\, {\rm eV}/c^2$. This
result can be compared with the exact value from Eq. (G12) of \cite{mcmh}.

\subsection{The core mass -- halo mass relation without central BH}
\label{sec_fmcmh}

Substituting the core mass-radius relation (\ref{fmrr2}) into Eq.
(\ref{mcmh2}), we obtain the core mass -- halo mass relation without central BH
\begin{equation}
\label{fmcmh1}
\frac{M_c}{(M_h)_{\rm min}}=\left ( \frac{M_h}{(M_h)_{\rm min}}\right
)^{3/8}.
\end{equation}
We recover the scaling from Eq.
(169) of \cite{mcmh}. Returning to the
original variables, we get 
\begin{equation}
\label{fmcmh2}
M_c=\frac{(2\pi)^{3/2}}{\alpha^{3/8}\nu^{3/4}}\left
(\frac{3\zeta}{20}\right
)^{3/4}\left (\frac{3}{\pi}\right )^{1/2}\left
(\frac{\hbar^{4}\Sigma_0M_h}{G^2m^{16/3}}\right )^{3/8}.
\end{equation}
The prefactor is $1.155$. For a DM halo of mass
$M_h=10^{12}\,
M_{\odot}$ similar to the one that surrounds our Galaxy, we obtain a core
mass $M_c=3.16\times 10^9\, M_{\odot}$  (we have
taken $(M_h)_{\rm min}=10^{8}\,
M_{\odot}$). The corresponding core radius is $R_c=201\, {\rm pc}$ [see Eq.
(\ref{fmrr2})]. The quantum core represents a bulge or a nucleus
(it cannot mimic a BH).

\subsection{The core mass -- halo mass relation with a central BH}
\label{sec_fmcmhbh}

We now take into account the presence of a central BH. Substituting the core
mass-radius relation (\ref{fmrr1}) into Eq.
(\ref{mcmh2}), we obtain the equation
\begin{equation}
\label{fmcmh3}
\left \lbrack \frac{M_c}{(M_h)_{\rm min}}+\frac{\lambda}{\nu}\frac{M_{\rm
BH}}{(M_h)_{\rm min}}\right \rbrack\left (\frac{M_c}{(M_h)_{\rm min}}\right
)^{1/3}=\left (\frac{M_h}{(M_h)_{\rm min}}\right )^{1/2}
\end{equation}
determining the core mass $M_c$ as a
function of the
halo mass $M_h$ in the
presence of a central BH of mass $M_{\rm BH}(M_h)$ given by Eq.
(\ref{bhmhm6}). 
We have introduced the mass scale from Eq. (\ref{fmrr6}) corresponding to the
minimum halo mass without central BH. Setting $M_c=M_h$ in Eq.
(\ref{fmcmh3}) we obtain the minimum halo mass
$(M_h)_{\rm min}(M_{\rm BH})$ in the presence of a central BH.
However, we have already explained in Sec. \ref{sec_mh} that we can neglect the
effect of the BH on
the minimum halo. The effect of the BH will be important only for larger halos
(see below). Therefore, the minimum halo mass in the presence of a central BH is
still given in very good approximation by Eq. (\ref{fmrr6}).

\begin{figure}[!h]
\begin{center}
\includegraphics[clip,scale=0.3]{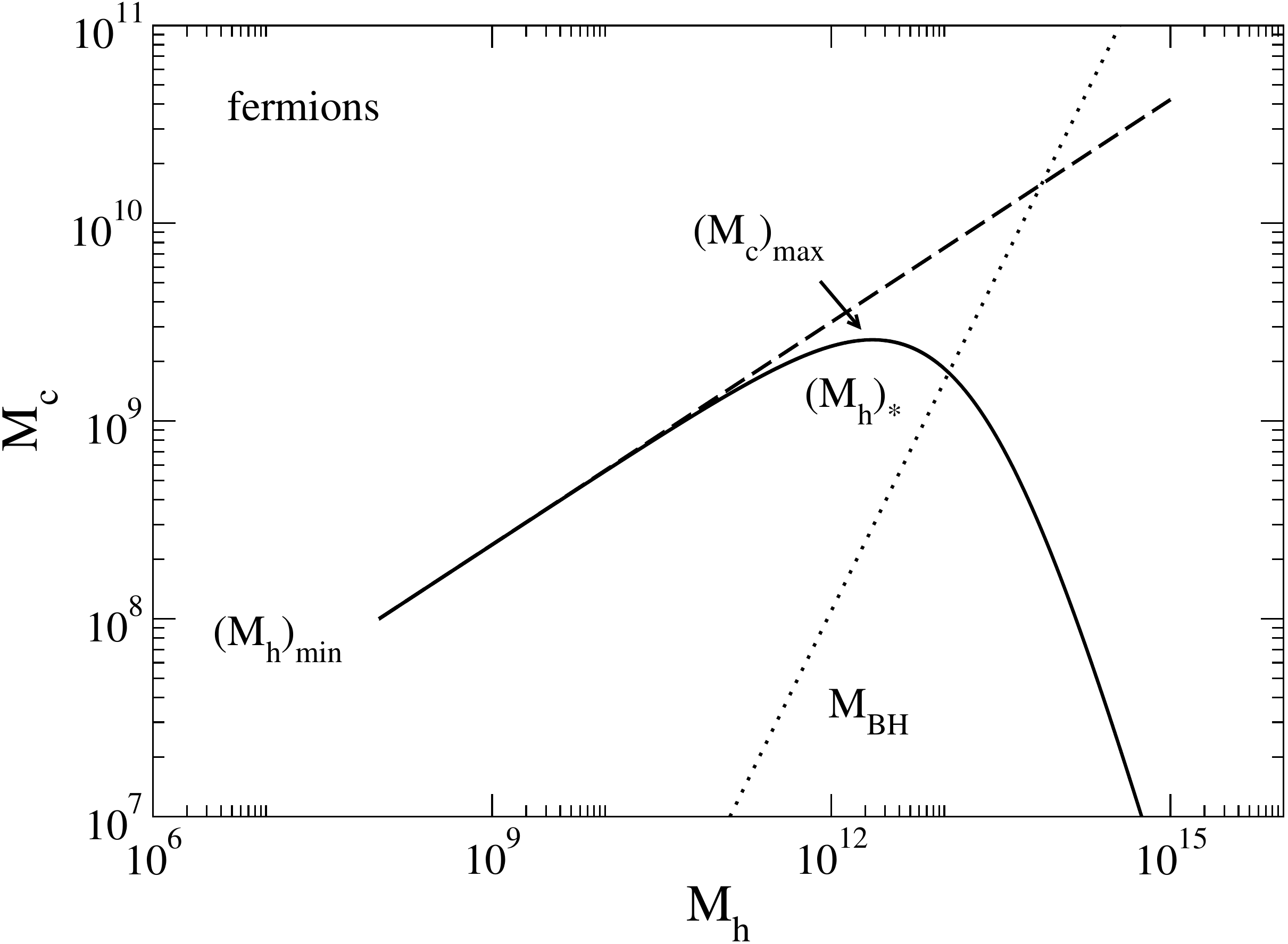}
\caption{Core mass $M_c$ as a function of the halo mass $M_h$ (solid line) for
fermions ($m=111\, {\rm eV}/c^2$). The mass is
normalized by $M_{\odot}$.  The function $M_c(M_h)$ presents a
maximum core mass $(M_c)_{\rm max,F}=2.57\times 10^{9}\,
M_{\odot}$ at $(M_h)_{\rm *,F}=2.31\times 10^{12}\, M_{\odot}$. All the
configurations are
stable. We have also represented the
relation $M_c(M_h)$ without BH (dashed line) and the
relation $M_{\rm BH}(M_h)$ (dotted line).}
\label{mhmcFERMI}
\end{center}
\end{figure}

The core mass -- halo mass relation (\ref{fmcmh3}) is plotted in Fig.
\ref{mhmcFERMI}. There is a maximum
core mass 
\begin{eqnarray}
\label{fmcmh4}
(M_c)_{\rm max, F}=2.57\times 10^9\, M_{\odot},
\end{eqnarray}
corresponding to a halo mass
\begin{eqnarray}
\label{fmcmh5}
(M_h)_{\rm *, F}=2.31\times 10^{12}\, M_{\odot}
\end{eqnarray}
and a BH mass
\begin{eqnarray}
\label{fmcmh6}
(M_{\rm BH})_{\rm *, F}=2.89\times 10^8\, M_{\odot}.
\end{eqnarray}
The effect of the BH becomes
important when  $M_{\rm BH}\sim M_c$, corresponding  to $(M_h)_{\rm *, F}\sim
10^{12}\, M_{\odot}$. When
$M_h\ll (M_h)_{*}$ the effect of the BH is
negligible and we recover the scaling 
\begin{equation}
\label{fmcmh7}
\frac{M_c}{(M_h)_{\rm min}}\sim\left ( \frac{M_h}{(M_h)_{\rm min}}\right
)^{3/8}
\end{equation}
corresponding to Eq. (\ref{fmcmh1}).

When $M_h\gg (M_h)_{*}$ the BH dominates and we get
the
scaling
\begin{eqnarray}
\label{fmcmh8}
\frac{M_c}{(M_h)_{\rm
min}}&\sim&\left (\frac{\nu}{\lambda}\right )^3 \left (\frac{(M_h)_{\rm
min}}{M_{\rm
BH}}\right )^3 \left (\frac{M_h}{(M_h)_{\rm
min}}\right
)^{3/2}\nonumber\\
&\sim& \left (\frac{\nu}{\lambda}\right )^3 \frac{1}{A^3} \left
(\frac{(M_h)_{\rm
min}}{M_{h}}\right )^{3a-3/2}.
\end{eqnarray}
This relation can be directly obtained from Eqs. (\ref{mcmh2})
and (\ref{fmrr1q}). It exhibits a critical index $a_{\rm F}=1/2$. When
$a>a_{\rm
F}$ the core mass
decreases with $M_h$ and when  $a<a_{\rm F}$ the core mass
increases with $M_h$. For the measured value $a=1.16$, we are in the first case.
For a DM halo of mass $M_h=10^{12}\,
M_{\odot}$ similar to the one that surrounds our Galaxy, we obtain a core
mass $M_c=2.39\times 10^9\, M_{\odot}$ a little smaller than the value
$M_c=3.16\times 10^9\, M_{\odot}$ 
obtained above in the
absence of a central BH (we have taken $(M_h)_{\rm min}=10^{8}\,
M_{\odot}$).

\subsection{Summary}
\label{sec_summaryf}

In the fermion case ($m=111\, {\rm eV}/c^2$ given by Eq. (\ref{fmrr7})) the
halos
with a mass $M_h\ge (M_h)_{\rm
min}$ [see Eq. (\ref{fmrr6})] contain a quantum core of mass $M_c$ given
by Eq.
(\ref{fmcmh3}). The core mass achieves a maximum value $(M_c)_{\rm max,F}$ at
$(M_h)_{\rm *,F}$
[see Eqs. (\ref{fmcmh4}) and (\ref{fmcmh5})]. All the configurations are
stable.

In this section, we have expressed the core mass -- halo mass relation
in terms of $M_h$. This relation  could be easily expressed in terms of $M_v$
by using Eq. (\ref{bhmhm2b}).

\section{Comparison with other works}
\label{sec_com}

Let us briefly compare our results with those obtained by Davies and Mocz
\cite{dm}.  These authors consider fuzzy DM
soliton cores made of
noninteracting bosons around
SMBHs and assume that the $M_c(M_h)$ relation is unchanged
by the presence of the SMBH. If we take into account the effect of the SMBH
in the manner performed in the present paper, we find that, for
noninteracting bosons, the effect of the BH becomes important for DM halos of
mass $(M_h)_{*,0}=1.96\times 10^{12}\, M_{\odot}$. Accordingly,
for the
DM halos
considered by Davies and Mocz
\cite{dm}, with mass $M_v=10^{12}\, M_{\odot}$ (Milky Way) and
$M_v=2\times 10^{14}\, M_{\odot}$ (elliptical galaxy in the Virgo cluster),
the presence of the SMBH is expected to change the value of the core mass
(especially in the second case),
except if it forms after the quantum core as assumed by these
authors.\footnote{As discussed by Davies and Mocz
\cite{dm} at the end of
their Sec. 4, their conclusions are robust and should not significantly change
if we properly account
for the modification of the quantum core mass due to the presence of the SMBH
as suggested above.
We stress, however, that their results (and all the results
quoted in their Fig. 1) assume that the BECDM particle is noninteracting. As
emphasized in our series of papers
\cite{prd1,modeldm,suarezchavanis3,mcmh}, and recalled in the
Introduction,
the
consideration of a self-interaction can considerably change the value of the DM
particle mass, up to $20$ orders of magnitude, and solve some tensions with
observations. The possibility of a self-interaction should be taken into
account  in
the literature when considering astrophysical constraints on the BECDM particle
mass.} The
same conclusion is reached in the case of fermions for which $(M_h)_{\rm
*,F}=2.31\times 10^{12}\, M_{\odot}$.
By contrast, for self-interacting
bosons in the TF limit, we find that the effect of the   BH becomes important
for DM halos of mass $(M_h)_{\rm *,TF}=6.92\times 10^{14}\, M_{\odot}$. This
mass scale is of the order of the mass of the largest DM halos in the Universe.
As a result, for most DM halos made of self-interacting
bosons in the TF limit, the effect of the  SMBH is negligible, or
only marginally important. Finally, for bosons with a repulsive
self-interaction, the effect of the BH is negligible when $a_s$ is close to
$(a_s)_{c}$ while it becomes important for halos of size $M_h \sim 10^{12}\,
M_{\odot}$ when $a_s$ is close to zero. From Fig. \ref{mhmcATT} and Table II
(which can be compared with Table I of \cite{mcmh}) we see that the transition
occurs for $a_s/a'_*\sim -0.00746$ (with $a'_*=4.95\times 10^{-62}\, {\rm
fm}$).

\section{Conclusion}
\label{sec_con}

In this paper, we have analytically derived the core mass -- halo mass
relation of  bosonic and fermionic DM halos in the presence of a central BH.
Our
results are summarized in Secs. \ref{sec_summary} (for bosons)
and \ref{sec_summaryf} (for fermions). They generalize the results
previously reported in \cite{mcmh} in
the case without central BH. To obtain our results, we have used the BH mass --
halo
mass relation (\ref{bhmhm1}) obtained from the observations \cite{bandara}, the
core
mass -- radius relations (\ref{mrr1}) and (\ref{fmrr1}) in the presence of a
central BH obtained
from a Gaussian ansatz in \cite{epjpbh}, and the velocity dispersion tracing
relation (\ref{mcmh1}) introduced in \cite{mocz,bbbs,modeldm} and justified from
an effective thermodynamic approach in \cite{modeldm,mcmh}. We have assumed that
this relation remains valid in the presence of a central BH.  We have also
assumed that the BH exists prior to the quantum core and that it affects its
mass-radius relation in the manner discussed in  \cite{epjpbh}. However, very
little is known about the mechanism of formation of SMBHs at the center of
galaxies. It could well be that the SMBH is formed after the quantum core, and
could even result from its gravitational collapse. In this respect,
we have argued in \cite{modeldm,mcmh} that above a critical halo mass
$(M_h)_{\rm MCP}$, related to a microcanonical critical point in our
effective thermodynamic model, the core-halo structure of DM halos becomes
unstable. In that case, the system undergoes a gravothermal catastrophe followed
by a dynamical instability of general relativistic origin \cite{balberg} leading
to the formation of a SMBH. This could be a manner to form SMBHs at the
center of galaxies, although our arguments suggest that this scenario
works only for sufficiently large halos ($M_h\gtrsim 10^{11}\,
M_{\odot}$). [Actually, the fact that SMBHs
can form only in sufficiently large galaxies is consistent with the
conclusion reached by
Ferrarese \cite{ferrarese} on the basis of observations.] Therefore,
different
mechanisms of SMBH formation at the center of galaxies are possible. They should
be investigated in future works.

\appendix

\section{The case of a fixed BH mass}

In main text, when studying the core mass -- halo mass
relation $M_c(M_h)$, we have taken into account the dependence of the BH mass
with the halo mass [see Eq. (\ref{bhmhm6})]. This is the astrophysically
relevant situation. However,  it is interesting to consider, for comparison, the
behavior of the  core mass -- halo mass relation $M_c(M_h)$ in the academic case
where the BH mass is fixed. This relation is determined by Eq.
(\ref{mcmh3}) or Eq. (\ref{mcmh6}), where $M_{\rm BH}$ is now a constant.  When
$a_s\ge 0$, the function $M_c(M_h)$ is monotonic.
When $a_s<0$, the core mass is maximum at
\begin{equation}
\label{}
\frac{(M_h)^{\rm fixed BH}_{\rm max}}{(M_h)_{\rm min,0}}=\frac{1}{4}\left
(\frac{a_*}{a_s}\right )^2
\end{equation}
as in the absence of a central BH \cite{mcmh}. It has the the value
\begin{eqnarray}
\label{}
\frac{(M_c)^{\rm fixed BH}_{\rm max}}{(M_h)_{\rm
min,0}}&=&-\frac{\lambda}{2\nu}\frac{M_{\rm
BH}}{(M_h)_{\rm
min,0}}\nonumber\\
&+&\sqrt{ \frac{\lambda^2}{4\nu^2}\left (\frac{M_{\rm BH}}{(M_h)_{\rm
min,0}}\right )^2
+\frac{1}{4}\frac{a_*}{|a_s|} }.
\end{eqnarray}
This expression coincides with the maximum core mass $(M_c)_{\rm max,*}$ given
by Eq. (\ref{att3}) [see also Eqs. (\ref{mrr8}) and (\ref{att3b})].

\end{document}